\documentclass[prd,aastex,amsmath,amssymb,twocolumn,widetext,floatfix,aps,showpacs]{revtex4-2} 

\usepackage{graphicx}% Include figure files
\usepackage{textcomp}
\usepackage{xcolor}
\definecolor{myblue}{rgb}{0.0, 0.0, 0.6}
\usepackage{hyperref}
\hypersetup{
  colorlinks = true,
  citecolor  = myblue,
  linkcolor  = myblue,
  urlcolor   = myblue
}
\begin{document}

\title{
  Breakup corrections to the ${}^3$He beam polarization measurements at the future BNL Electron Ion Collider
}%

\author{A.~A.~Poblaguev}\email{poblaguev@bnl.gov}

\affiliation{%
 Brookhaven National Laboratory, Upton, New York 11973, USA
}%
\date{December 12, 2022}

\begin{abstract}
  Requirements for hadron polarimetry at the future Electron-Ion Collider (EIC) include measurements of absolute ${}^3\text{He}\,(h)$ beam polarization with systematic uncertainties better than $\sigma^\text{syst}_P/P\lesssim1\%$. Due to the successful use, since 2005, of the polarized hydrogen jet target polarimeter (HJET) to measure proton beam polarization at the Relativistic Heavy Ion Collider (RHIC), the HJET technique promises to be suitable for the ${}^3\text{He}$ polarimetry at EIC. For that, however, one needs to know the ratio of the $h^\uparrow{p}$ and $p^\uparrow{h}$ analyzing powers. For elastic scattering, this ratio can be evaluated with sufficient accuracy if $p^\uparrow{p}$ analyzing power is precisely known. In this paper, the deuteron beam data acquired at HJET in RHIC Run\,16 was used to evaluate corrections to the measured helion beam polarization due to the ${}^3\text{He}$ breakup. The breakup effect was found to be negligible if the ratio of the beam and target (jet) single spin asymmetries are concurrently measured to determine the  ${}^3\text{He}$ beam polarization.
\end{abstract}

\maketitle

\section{Introduction}

High energy ($\sim\!100\,\text{GeV/nucleon}$) polarized ($\gtrsim\!70\%$) helion ($h$), i.e., ${}^3\text{He}$ ($A_h\!=\!3$, $Z_h\!=\!2$), beams are planned for the future Electron-Ion Collider (EIC) \cite{Accardi:2012qut}. The EIC physics program requirement for the beam polarization measurement accuracy is \cite{AbdulKhalek:2021gbh}
\begin{equation}
  \sigma_P^\text{syst}/P \lesssim 1\%.
  \label{eq:systEIC}
\end{equation}

The development of the helion beam polarimetry for EIC is greatly influenced by successful operation of the Atomic Polarized Hydrogen Gas Jet Target (HJET)\,\cite{Zelenski:2005mz} at the Relativistic Heavy Ion Collider (RHIC). For elastic scattering of the vertically polarized RHIC proton beam off the HJET proton target (the jet), both the beam $a_\text{beam}\!=\!A_\text{N}P_\text{beam}$ and target $a_\text{jet}\!=A_\text{N}P_\text{jet}$ spin-correlated asymmetries were concurrently measured by counting the recoil protons in the left/right symmetric silicon strip detectors\,\cite{Poblaguev:2020qbw}. Since the jet polarization is well determined, e.g., $P_\text{jet}\!=\!0.957\pm0.001$ in RHIC Run\,17, the beam polarization can be readily derived from the measured asymmetries:
\begin{equation}
  P_\text{beam}=P_\text{jet}\times a_\text{beam}/a_\text{jet}.
  \label{eq:PbeamHJET}
\end{equation}
Systematic error in the measurements was evaluated\,\cite{Poblaguev:2020qbw} as  $\sigma_P^\text{syst}/P \lesssim 0.5\%$.

For the elastic scattering, analyzing power $A_\text{N}(s,t)$ is, generally, a function of the center-of-mass energy squared $s\!=\!2m_pE_\text{beam}$ and the momentum transfer squared $t\!=\!-2m_pT_R$, where $m_p$ is a proton mass, $E_\text{beam}$ is the beam energy, and $T_R$ is the recoil proton kinetic energy. For the Coulomb-nuclear interference (CNI) elastic proton-proton scattering, the commonly used parametrization of $A_\text{N}(t)$ was introduced in Ref.\,\cite{Buttimore:1998rj}. Some small corrections to $A_\text{N}(t)$ which were neglected in Ref.\,\cite{Buttimore:1998rj} but which appeared to be noticeable in the HJET data analysis were discussed in Ref.\,\cite{Poblaguev:2019vho}. In RHIC Runs\,15 $(E_\text{beam}\!=\!100\,\text{GeV})$ and 17 (255\,GeV), the elastic ${pp}$ analyzing power
\begin{equation}
  A_\text{N}^{pp}(t) = a_\text{jet}(T_R)/P_\text{jet}
\end{equation}
was determined\,\cite{Poblaguev:2019saw} with accuracy $|\delta A_\text{N}^{pp}(t)|\!\approx\!0.002$ in the $0.0013\!<\!-t\!<\!0.018\,\text{GeV}^2$ momentum transfer range. Also, the hadronic spin-flip amplitude parameter\,\cite{Buttimore:1998rj},
\begin{equation}
  r_5 = R_5 + iI_5,\qquad|r_5|\approx0.02,
\end{equation}
was reliably isolated for both beam energies. The measured elastic $p^\uparrow{p}$ values of $r_5$ can be used to evaluate, with sufficient accuracy, the single spin-flip amplitudes in elastic $p^\uparrow{h}$ and $h^\uparrow{p}$ scattering\,\cite{Poblaguev:2022yzw}. 

Although HJET was designed to measure proton beam polarization, beginning in 2015 it also routinely operated in the RHIC (unpolarized) ion beams ${}^2\rm{H}^{+}~(d)$, ${}^{16}O^{8+}$, ${}^{27}\text{Al}^{12+}$, ${}^{96}\text{Zr}^{40+}$, ${}^{96}\text{Ru}^{44+}$, ${}^{197}\text{Au}^{79+}$. The main purpose was a study of systematic errors (at HJET) and measurements of the $p^\uparrow{A}$ analyzing power. It was found that for ion beams the recoil proton spectrometer performance is nearly the same as that for a proton beam.

To apply the HJET technique to the ${}^3\text{He}^\uparrow$ beam polarimetry, one should know the ratio of the $h^\uparrow{p}$ and $p^\uparrow{h}$ analyzing powers. In leading order approximation\,\cite{Kopeliovich:1974ee} for the analyzing power, Eq.\,(\ref{eq:PbeamHJET}) should be rewritten as\,\cite{Buttimore:2009zz}
\begin{align}
    P_\text{beam}^h &= P_\text{jet}\,a_\text{beam}^h/a_\text{jet}^p
  \nonumber \\
  &\times \frac{\mu_p-1}{\mu_h/Z_h-m_p/m_h}\left[1+\text{corr}\right],
  \label{eq:PolBeamR} 
\end{align}
where $\mu_h\!=\!-2.128$ and $\mu_p\!=\!2.793$ are magnetic moments of a helion and a proton, respectively. For the elastic scattering, a few percent correction due to hadronic spin-flip amplitudes can be calculated\,\cite{Kopeliovich:2000kz,Poblaguev:2022yzw} if the proton-proton value of $r_5^{pp}$ (for the same beam energy per nucleon) is pre-determined. The main goal of this paper is to evaluate possible systematic uncertainties due to elastic data contamination by inelastic (breakup) events $h\!\to\!pd$.

The breakup rate is expected to be very small compared to the elastic one in CNI region\,\cite{Igo:2003cs}. In the experiment\,\cite{Bellettini:1966zz}, the angular distributions of $\approx\!20\,\text{GeV}$ protons scattered by eight nuclei (from ${}^6\rm{Li}$ to ${}^{238}\rm{U}$) were measured in the scattering angle range equivalent to $0.0016\!<\!-t\!<\!0.16\,\text{GeV}^2$. The events in which pion production took place were rejected by the event selection cuts. The breakup and elastic contributions to the measured $d\sigma/d\Omega$ were separated in the Glauber theory based analysis\,\cite{Glauber:1970jm}. It was found that for all studied ions, the breakup/elastic rate ratio is $\lesssim\!0.1$ in the momentum transfer range $0.002\!<\!-t\!<\!0.02\,\text{GeV}^2$ used in HJET. However, the breakup rate dependence on the ion missing mass was not given there.

Here, to evaluate the breakup fraction for recoil protons detected in the helion scattering in HJET, I analyzed the deuteron beam data acquired in RHIC Run\,16\,\cite{Liu:IPAC2017-TUPVA046}. The obtained result was extrapolated to a helion beam scattering. At HJET $(T_R\!<\!10\,\text{MeV})$, the helion beam elastic data contamination by the ${}^3\text{He}$ breakup events was estimated to be small, $\lesssim\!0.5\%\!\times\!T_R/\text{MeV}$ and the consequent systematic corrections well cancel in expression (\ref{eq:PolBeamR}) for the measured beam polarization.

\section{\label{sec:inel}
  Inelastic scattering in HJET
}

\subsection{The recoil proton kinematics}

\begin{figure}[t]
  \begin{center}
    \includegraphics[width=0.9\columnwidth]{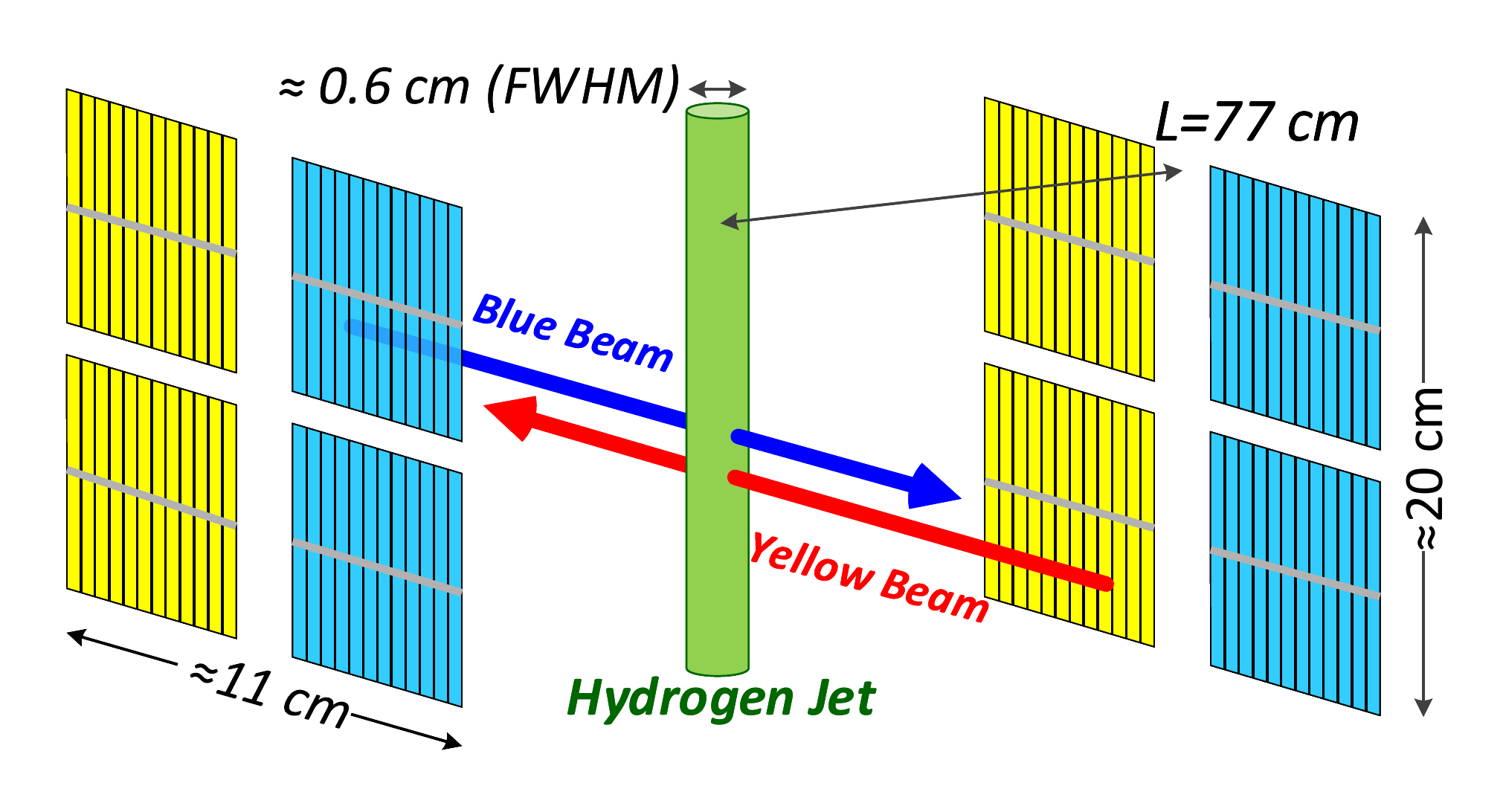}
  \end{center}
  \caption{\label{fig:HJET}
    Schematic view of the HJET recoil spectrometer. Each of the detectors consists of 12 vertically oriented Si strips.
  }
\end{figure}

Measurement at the HJET\,\cite{Poblaguev:2020qbw} is depicted in Fig.\,\ref{fig:HJET}. Both RHIC beams, so-called {\em blue} and {\em yellow}, cross the hydrogen jet target and the recoil protons are counted in silicon strip detectors.  Four detectors are designated to measure recoil protons from the {\em blue} beam and the other four from the {\em yellow} one. 

At HJET, since only the jet recoil protons $p_j$ are detected, an inclusive scattering is studied:
  \begin{equation}
    A+p_j \to X+p_j,\,\qquad \Delta=(M_X^2-m_A^2)/2m_A.
    \label{eq:Inelastic}
  \end{equation}
In Eq.\,(\ref{eq:Inelastic}), $A$ denotes a scattered beam ion, in particular a proton, and missing mass $M_X$ is parametrized by a mass excess $\Delta$.
  
  To isolate the elastic scattering ($\Delta\!=\!0$), the following dependence of the recoil proton $z$ coordinate (along the beam) in a detector on $T_R$ and $\Delta$ can be used:
\begin{equation}
  \frac{z-z_\text{jet}}{L} = \sqrt{\frac{T_R}{2m_p}}\times\left[1+\frac{m_p}{E_\text{beam}}\left(\frac{m_p}{m_A}+\frac{\Delta}{T_R}\right)\right]\!,
  \label{eq:tanTheta}
\end{equation}
where $L\!=\!77\,\text{cm}$ is the distance from the scattering point to the detector and $E_\text{beam}$ is the beam energy per nucleon. Possible values of the scattering point coordinate $z_\text{jet}$ are
 predetermined by the jet profile density: $\langle z_\text{jet}\rangle\!=0$ and $\langle z^2_\text{jet}\rangle^{1/2}\!=\!\sigma_\text{jet}\!\approx\!0.26\,\text{cm}$. In HJET measurements, $z$ coordinate in a Si detector is discriminated by the strip number $k=0,\dots,11$:
\begin{equation}
  z_k = 0.9+k\times0.375\:\text{cm}.
\end{equation}
For such defined $z_k$, the coordinate of a strip $k$ center is $z_\text{strip}^{(k)}=(z_k+z_{k+1})/2$.

For a fixed value of the recoil proton energy $T_R$, the elastic event rate has a maximum in strip $k_T$ defined by a condition
\begin{equation}
  z_\text{strip}^{(k_T)} \approx L\,\sqrt{T_R/2m_p}
  \approx 1.8\,\text{cm}\times\sqrt{T_R/\text{MeV}}.
\end{equation}
In the strip number units, the rms of the elastic peak is about $\sigma_k\!\approx\!0.7$. Since, according to Eq.\,(\ref{eq:tanTheta}), inelastic events can be detected only in the strips $k\!>\!k_T$, one can evaluate\,\cite{Poblaguev:2020qbw} the inelastic fraction by comparing background in strips $k\!>\!k_T\!+\!\nu\sigma_k$ and $k\!<\!k_T-\nu\sigma_k$ where a cutoff factor $\nu$ should be about 3--4 to eliminate the elastic data.

The inelastic recoil protons cannot be detected in HJET if
\begin{equation}
  \Delta/E_\text{beam} > z_{12}^2/2L^2
  \approx 2.5\times10^{-3}.
  \label{eq:DeltaMax}
\end{equation}
Here, I neglect the term $m_p^2/m_AE_\text{beam}$ in Eq.\,(\ref{eq:tanTheta}) and do not consider smearing due to the jet thickness $\sigma_\text{jet}$ and due to the alignment of the detectors. For small values of $\Delta$, the inelastic events cannot be separated from the elastic ones if
\begin{equation}
  \Delta/E_\text{beam} < \frac{\nu\sigma_\text{jet}}{L}\sqrt{\frac{2T_R}{m_p}}
    \approx 0.9\times10^{-3}.
  \label{eq:DeltaMin}
\end{equation}
The numerical estimate was given assuming $T_R\!=\!4\,\text{MeV}$ and $\nu\!=\!3$ standard deviation cutoff to isolate the elastic events. 

For an incident ion $A$, the following inelastic processes can, generally, be considered:\\
$\bullet$\;%
{\em meson production: $A\to \pi+X$.}
The method used to eliminate such events from the elastic data in the proton beam polarization measurements at HJET\,\cite{Poblaguev:2020qbw} is also applicable for the ${}^3\rm{He}$ beam at EIC. Therefore, such an inelastic scattering is not analyzed in this paper.\\
$\bullet$\;%
{\em beam ion breakup: $A\to A_1+A_2+\dots$.}
For the ${}^3\rm{He}$ beam, two such processes are possible: $h\!\to\!pd$ $(\Delta_\text{thr}\!=\!5.5\,\text{MeV})$ and $h\!\to\!ppn$ $(\Delta_\text{thr}\!=\!7.7\,\text{MeV})$. Due to phase space suppression factors at low $\Delta$, I will consider only two-body breakup $h\!\to\!pd$.\\
$\bullet$\;%
{\em beam ion excitation: $A \to A^*$.}
There are no proven excited states for ${}^3\rm{He}$ \cite{Purcell:2015gtm}.\\
$\bullet$\;%
{\em Inner Bremsstrahlung (IB): $A \to A+\gamma$.}
The IB fraction can be readily evaluated in the soft photon approximation\,\cite{Berestetsky:1982aq} giving a negligible value, $d\sigma_\text{IB}/d\sigma_\text{el}\!=\!|\alpha{Z}^2t/\pi m_A^2|\,\ln{\Delta_\text{max}/\Delta_\text{min}}$, for the HJET measurements. Here $Z$ and $A$ are atomic and mass numbers of the beam ion. The considered range of the scattered mass excess is defined by $\Delta_\text{min}$ and $\Delta_\text{max}$.

\subsection{The proton beam inelastic scattering}

For inelastic $\mathit{pp}$ scattering, $\Delta\!\ge\!m_\pi\!=\!135\,\text{MeV}$. Therefore see Eq.\,(\ref{eq:DeltaMax}), the inelastic recoil protons can hit the HJET detectors only if the proton beam energy is $E_\text{beam}\!>\!55\,\text{GeV}$.

Nonetheless, in RHIC polarized proton Run\,15 $(E_\text{beam}\!=\!100\,\text{GeV})$ the elastic data contamination by {\em meson production} events was found\,\cite{Poblaguev:2020qbw} to be negligible (although the inelastic events can be clearly observed in Si strips numbered $k\!=\!11$). 

For the $255\,\text{GeV}$ beam (Run\,17), the {\em meson production} fraction in the detected events was about 5--10\%, which  required an adjustment of the background subtraction method and the elastic event selection cuts\,\cite{Poblaguev:2020qbw}.

\subsection{The Au beam measurements at HJET}

In gold beam measurements at HJET, the beam energy was mostly within a range of 3.85--31.3\,GeV/nucleon. Therefore, only Au {\em excitation} ($\Delta\!<\!3.6\,\text{MeV}$ \cite{HUANG2005283}) and Au {\em breakup} (the proton and neutron extraction energies are 5.78 and 8.07 MeV, respectively) inelastic events could be potentially found in the acquired data. Since no evidence of such events was experimentally found in the excess mass range of $4\!\lesssim\!\Delta\!\lesssim\!80\,\text{MeV}$, the following constraint on the quasielastic (breakup) fraction in the elastic data can be guesstimated:
\begin{equation}
  \sigma_\text{qel}^{p\text{Au}}/\sigma_\text{el}^{p\text{Au}}<\text{few}\times10^{-3}.
\end{equation}

Recently, in Run\,21, RHIC was filled, for special studies, with a single ({\it blue}) Au beam.  For two beam energies, 3.85 and 26.5 GeV/nucleon, HJET measurements were done with the holding field magnet switched off. Although the statistics accumulated were relatively low, the reduced background uncertainties in these measurements allowed more definite evaluation of the breakup fraction more. For the recoil proton energy range $1.3\!<\!\sqrt{T_R}\!<\!2.1\,\text{MeV}^{1/2}$ $(0.003\!<\!-t\!<\!0.009\,\text{GeV}^2)$, the following preliminary estimates of $\langle\sigma_\text{qel}^{p\text{Au}}/\sigma_\text{el}^{p\text{Au}}\rangle$, averaged over $T_R$ and $\Delta$, were done\,\cite{Au_Breakup}:
\begin{align}
  \!\!3.85\,\text{GeV}\!:\; &\quad0.20\pm0.12\,\% &\![3.6\!<\!\Delta\!<\!8.5\,\text{MeV}], \\
  \!\!26.5\,\text{GeV}\!:\; &-\!0.08\pm0.06\,\% &[20\!<\!\Delta\!<\!60\,\text{MeV}]. 
\end{align}

\subsection{A model for the beam ion breakup}

The observed suppression of the breakup events can be readily explained by an assumption that such a process is dominated by incoherent scattering of the jet proton (in the ion system) off a nucleon in the ion. If $p_x$ is an internal motion momentum of the nucleon (in the direction to the detector) then Eq.\,(\ref{eq:tanTheta}) leads to  
\begin{equation}
  \Delta = \left(1-\frac{m_p}{m_A}\right)T_R+p_x\sqrt{\frac{2T_R}{m_p}}.
  \label{eq:Delta}
\end{equation}
Assuming, for an estimate, $|p_x|\!<\!250\,\text{MeV}/c$ and $T_R\!<\!10\,\text{MeV}$, one finds $\Delta\!<\!50\,\text{MeV}$. Thus, in the considered case, the breakup fraction should be strongly suppressed by a phase space factor. In other words, the HJET recoil spectrometer design efficiently prevents the detecting of recoil protons from the beam ion breakup scattering.

To search for breakup events in HJET data, the following simplified model was used to parametrize $d^2N/dT_Rd\Delta$ event rate.

If a nucleon momentum distribution is given by $f(p_x,\sigma)\,dp_x$, one can expect the following event rate dependence on $\Delta$:
\begin{equation}
  dN/d\Delta \propto \tilde{f}(t,\Delta) = f(\Delta - \Delta_0,\sigma_\Delta)\times\Phi(\Delta),
  \label{eq:dNdD}
\end{equation}
where the distribution width parameter $\sigma$ is explained in the Appendix\,\ref{sec:func}, $\Delta_0\!=\!(1\!-\!m_p/m_A)T_R$, $\sigma^2_\Delta\!=\!2\sigma^2T_R/m_p$, and $\Phi(\Delta)$ is the phase space integral.

Generally, elastic scattering cross section can be described as
\begin{equation}
  d\sigma_\text{el} \propto |\phi_\text{el}|^2\times d\Phi_2(p_b+p_j;p_h,p_R),
\end{equation}
where $p_b$, $p_j$, $p_R$, and $p_h$ are four-momentum of the beam ${}^3\text{He}$, target (jet) proton, recoil proton, and scattered ${}^3\text{He}$, respectively. Assuming possible breakup of the helion, in the phase space factor,
\begin{equation}
  d\Phi_2(P;p_h,p_R)= \delta^4(P\!-\!p_h\!-\!p_R)\,%
  \frac{d^3p_h}{(2\pi)^32E_h}\,\frac{d^3p_R}{(2\pi)^32E_R},
\end{equation}
it is convenient to replace the ${}^3\text{He}$ term by
\begin{equation}
  \frac{d^3p_h}{(2\pi)^32E_h}=\delta(p_h^2-q^2)\frac{d^4p_h}{(2\pi)^3}
  \times dq^2\delta(q^2-m_h^2).
  \label{eq:dph}
\end{equation}

Considering elastic and breakup scattering only and discriminating the final state by the recoil proton energy and angle (i.e., by Lorentz invariants $t$ and $\Delta$), one can effectively substitute the elastic amplitude by a sum,
\begin{align}
  \phi_\text{el}(t) &\to \phi_\text{el}(t)+\sum_i{\int{\phi_i(t,\Delta)}\,d\Delta}
  \nonumber \\
  &= \phi_\text{el}(t)\times\left[1+\int{\psi(t,\Delta)\,d\Delta}\right],
  \label{eq:phi_brk}
\end{align}
where $i$ enumerates the breakup channels, $\psi(t,\Delta)\!=\!\phi_i(t,\Delta)/\phi_\text{el}(t)$ and there is no mutual interference between amplitudes discriminated by values of $\Delta$ and/or $i$. For sake of simplicity, only one breakup channel was considered and, thus, index $i$ was omitted in Eq.\,(\ref{eq:phi_brk}).

So, for the ${}^3\text{He}$ breakup, 
\begin{equation}
  p + h \to p + (p+d)_{h^*},\quad\Delta>\Delta^h_\text{thr}=5.5\,\text{MeV},
\end{equation}
one can substitute in Eq.\,(\ref{eq:dph})
\begin{equation}
  \!\!dq^2\delta(q^2-m_h^2) \to d\Delta\,\tilde{f}(t,\Delta) |\psi(t,\Delta)|^2 d\Phi_2(q;p_d,p_p).
\end{equation}
In this paper, to fit experimental data, $\psi(t,\Delta)$ was replaced by an average value $|\psi|\!=\!\langle|\psi(t,\Delta)|^2\rangle^{1/2}$ and approximate $\tilde{f}$ [Eq.\,(\ref{eq:dNdD})] by a momentum distribution function defined in the Appendix\,\ref{sec:func} with width $\sigma$ being a free parameter in the fit. 

Integrating over $\Delta$, one can relate the quasi-elastic (breakup) cross section to the elastic one:
\begin{align}
  \frac{d\sigma_\text{qel}}{d\sigma_\text{el}} =&~\omega(t) = |\psi|^2\,\omega_\Phi(t);%
  \label{eq:omega}\\
  \omega_\Phi(t)=&~\frac{\sqrt{2m_pm_d}}{4\pi m_h}\!\times\!%
  \int_{\Delta^h_\text{thr}}^{\infty}{d\Delta\,\tilde{f}(t,\Delta)\sqrt{ \frac{\Delta\!-\!\Delta^h_\text{thr}}{m_h} } }.
  \label{eq:phase}
\end{align}

Function $\omega(t)$ is supposed to be derived from the experimental data analysis, while $\omega_\Phi(t)$ can be calculated for any $\tilde{f}(t,\Delta)$ used.
  To study spin-flip effects, it is helpful to introduce a function 
  \begin{equation}
    \tilde{\omega}(t) = |\psi|\,\omega_\Phi(t)=\sqrt{\omega(t)\omega_\Phi(t)}.
    \label{eq:Xomega}
  \end{equation}

For low $t$, one can expect $\omega(t)\!\propto\!t$. However, considering actual energy threshold in detecting recoil protons and possible effective alteration of $\omega(t)$ due to event selection cuts, the breakup rates were approximated by linear functions:
\begin{eqnarray}
  \omega(T_R) &=& \omega_0 + \omega_1T_R,
  \label{eq:omega-nf}   \\
  \tilde{\omega}(T_R) &=& \tilde{\omega}_0 + \tilde{\omega}_1T_R.
  \label{eq:omega-sf}
\end{eqnarray}

\section{\label{sec:dAu}
  Experimental evaluation of the deuteron beam  breakup rate at HJET
}

In RHIC Run\,16, a beam energy scan of deuteron--gold collision was done. The deuteron ({\em blue} beam) energies used were  9.9, 19.6, 31.3, and 100.7\,GeV/nucleon. Due to HJET operation in this non-polarized beam Run, the deuteron beam breakup fraction in the forward elastic scattering was experimentally evaluated.

As explained in Sec.\,\ref{sec:inel}, to separate the elastic and inelastic events in the experimental data, one can compare the measured values of recoil proton $z$-coordinate and kinetic energy $T_R$. However, to isolate the inelastic events, a \lq\lq{regular}\rq\rq background for elastic events must be subtracted first. The background subtraction method, used in the HJET data analysis, is described in detail in Ref.\,\cite{Poblaguev:2020qbw}. It is based on an expectation that the background has the same rate (for given recoil proton energy $T_R$) in all strips of a Si detector. Thus, following Eq.\,(\ref{eq:tanTheta}), the inelastic event rate can be found by the difference in the measured background rates in the Si strips with large and small numbers. However, some corrections due to the recoil proton tracking in the holding field magnet and due to detectors being shadowed by the RF shield\,\cite{Poblaguev:2020qbw} should be taken into account.

\begin{figure}[t]
  \begin{center}
    \includegraphics[width=0.85\columnwidth]{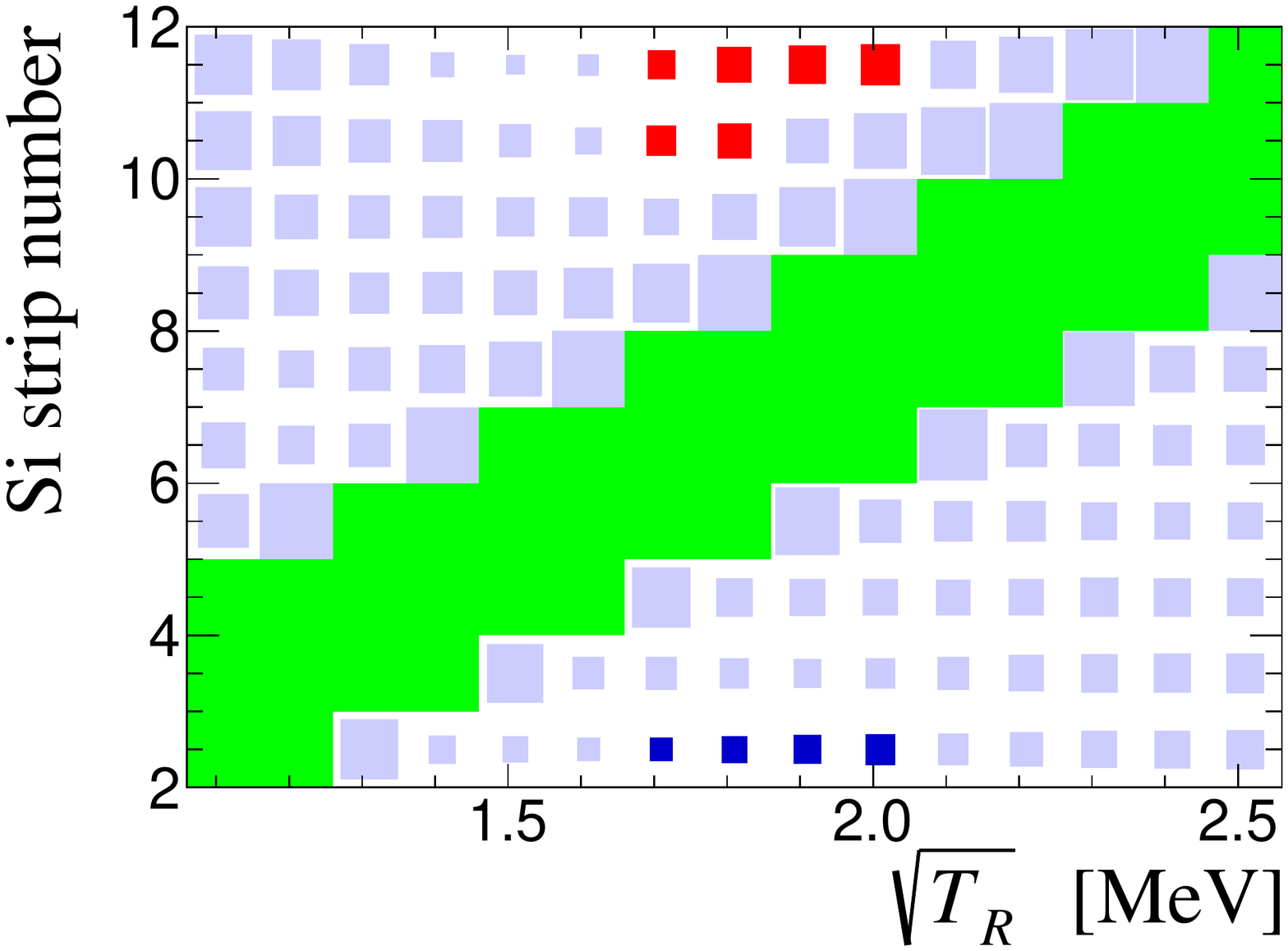}
  \end{center}
  \caption{\label{fig:10GeV}
    Recoil proton distribution in a {\em blue} Si detector for the 9.9\,GeV/nucleon deuteron beam. The elastic event statistics (green boxes) is up to $75\text{k/bin}$, but the displayed statistics (box size) is cut off at $5\text{k/bin}$. Red boxes in Si strips 10--11 (background + breakup) and blue boxes in strip 2 (background) were used to evaluate the breakup rate.
  }
\end{figure}

\begin{figure}[t]
  \begin{center}
    \includegraphics[width=0.85\columnwidth]{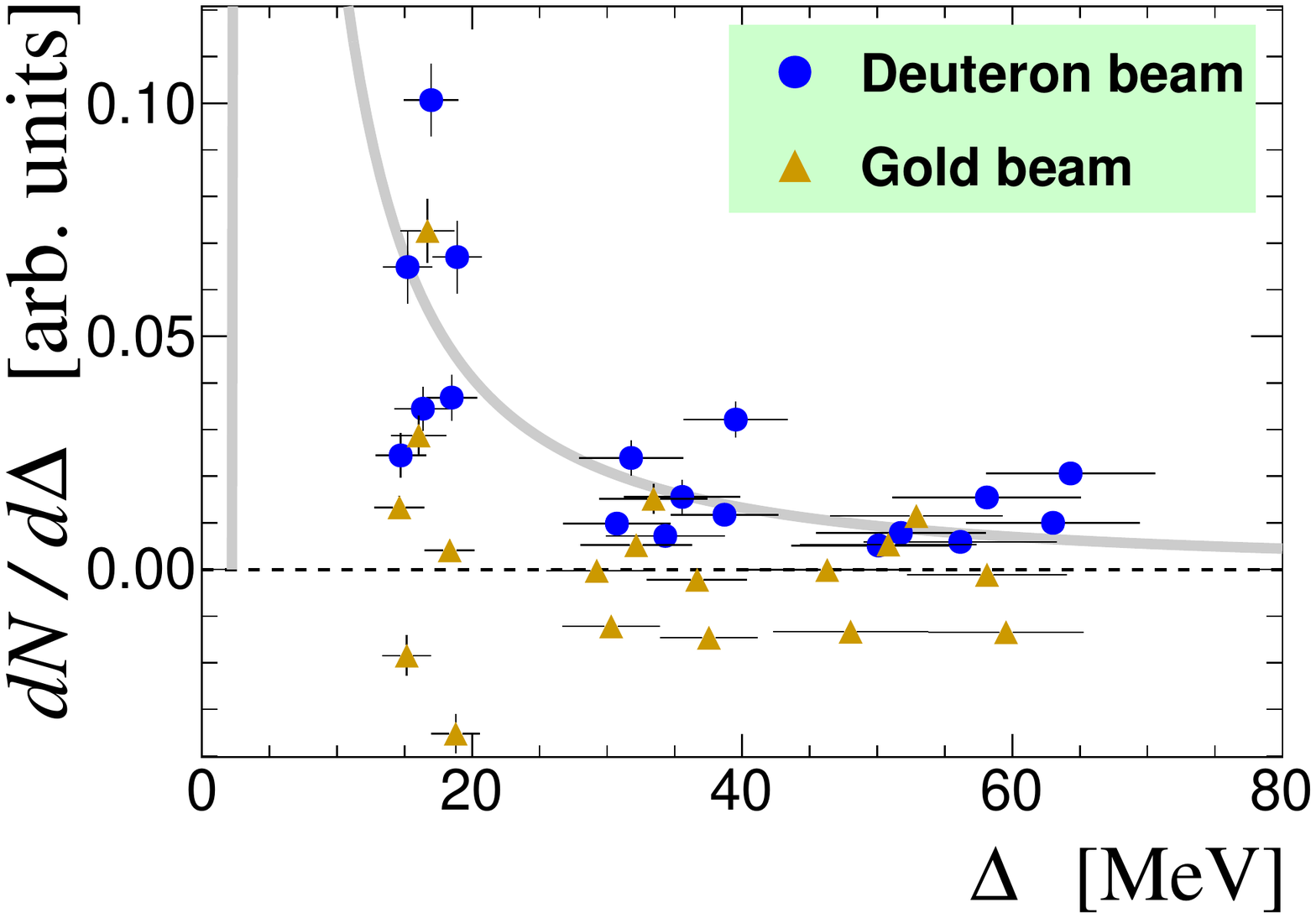}
  \end{center}
  \caption{\label{fig:Rate}
    The breakup rates (arbitrary units) in the selected data bins for deuteron and Au  beams. The solid line stands for a unity maximum (at $\Delta\!=\!3.1\,\text{MeV}$) normalized  $dN/d\Delta$ dependence which was calculated for $T_R\!=\!3.5\,\text{MeV}$ assuming $f_\text{BW}$ based distribution, $\sigma\!=\!35\,\text{MeV}$, $\Delta_\text{thr}\!=\!2.2\,\text{MeV}$, and $\psi(T_R,\Delta)\!=\!\text{const}$.}
\end{figure}

\begin{table*}[t]
  \begin{center}
    \begin{tabular}{l|r r| r r r | r}
      &          &          &\multicolumn{3}{c|}{\bf Deuteron Beam} & {\bf Gold Beam}
      \\[\smallskipamount]
      $f(p_x,\sigma)$~~ &
      \multicolumn{1}{c}{~$\sigma\,\text{[MeV/c]}$~} &
      \multicolumn{1}{c|}{$\chi^2$} &
      \multicolumn{1}{c}{$n_d$} &
      \multicolumn{1}{c}{$f_d$ [\%]} &
      \multicolumn{1}{c|}{$f_h$ [\%]}  &
      \multicolumn{1}{c}{$~~~n_\text{Au}$}
      \\[\smallskipamount]
      & & & &
      \multicolumn{1}{c}{~(2.8--4.2\,MeV)~} &
      \multicolumn{1}{c|}{~(1--10\,MeV)~}   &
      \\[\smallskipamount]
      \hline
      $f_\text{H} $ &  30~~~~ &~~403.1~~~&$~~0.923\pm0.198~~$&$11.1\pm2.4~~~$&$4.0\pm0.9~~$&$  0.135\pm0.189$\\
      $f_\text{H} $ & 122~~~~ &  143.6~~~&$  0.180\pm0.019~~$&$ 3.3\pm0.3~~~$&$1.9\pm0.2~~$&$~-0.010\pm0.029$\\
      $f_\text{G} $ & 240~~~~ &  231.9~~~&$  0.132\pm0.019~~$&$ 2.6\pm0.4~~~$&$1.6\pm0.2~~$&$ -0.007\pm0.022$\\
      $f_\text{BW}$ &  35~~~~ &   96.6~~~&$  0.400\pm0.033~~$&$ 5.7\pm0.5~~~$&$2.7\pm0.2~~$&$ -0.022\pm0.065$\\[\smallskipamount]
      \hline
      \multicolumn{3}{r|}{$1/\chi^2$ weighted average~~~}&  & $5.0\pm1.4~~~$  & $2.4\pm0.4~~$ & $0.000\pm0.027$ 
    \end{tabular}
  \end{center}
  \caption{\label{tab:dAu}
    Fit of the breakup component in the 10--31\,GeV/nucleon deuteron beam experimental data depending on the nucleon momentum distribution model. Statistical errors scaled by a factor $(\chi^2/\text{NDF})^{1/2}$ are shown. The simulation normalization factor $n_d$ is defined in Eq.\,(\ref{eq:norm}). The subsequent estimates of the elastic data contamination by breakup events for deuteron, $f_d$, and ${}^3\text{He}$, $f_h$, beams are given for the specified recoil proton energy ranges. $n_\text{Au}$ is the normalization factor obtained if the concurrently acquired gold beam data in a {\em yellow} beam detector were used in the analysis.  
  }
\end{table*}

Since HJET had operated in parasitic mode in Run\,16, the vertical beam positions in HJET were not optimized for the recoil proton detection. As result, the recoil protons were efficiently detected only in lower {\em blue} and upper {\em yellow} detectors. In addition, to suppress the magnetic field tracking effects\,\cite{Poblaguev:2020qbw}, only left (relative to the beam direction) detectors and only recoil protons with energy above $T_R^{1/2}\!>\!1.6\,\text{MeV}^{1/2}$ were used in the data analysis. For the 9.9\,GeV/nucleon deuteron beam, the event rate distribution in the lower left {\em blue}  detector is shown in Fig.\,\ref{fig:10GeV}. The bins in which elastic events are dominant, are colored green. The data in strips \#10 and \#11, marked red, were used to evaluate the breakup event rate. The background was evaluated using dark blue bins in strip \#2.

For the selected breakup events bins, a $\Delta$ range, suitable for the breakup fraction evaluation, depends on the beam energy: 12--21\,MeV (9.9\,GeV/nucleon), 25--44\,MeV (19.6\,GeV/nucleon), and 41--72\,MeV (31.3\,GeV/nucleon). The 100\,GeV/nucleon data were not used in the fit because they contain substantial (for the breakup rate study) contribution from {\em meson production} inelastic events.

\begin{figure}[t]
  \begin{center}
    \includegraphics[width=0.85\columnwidth]{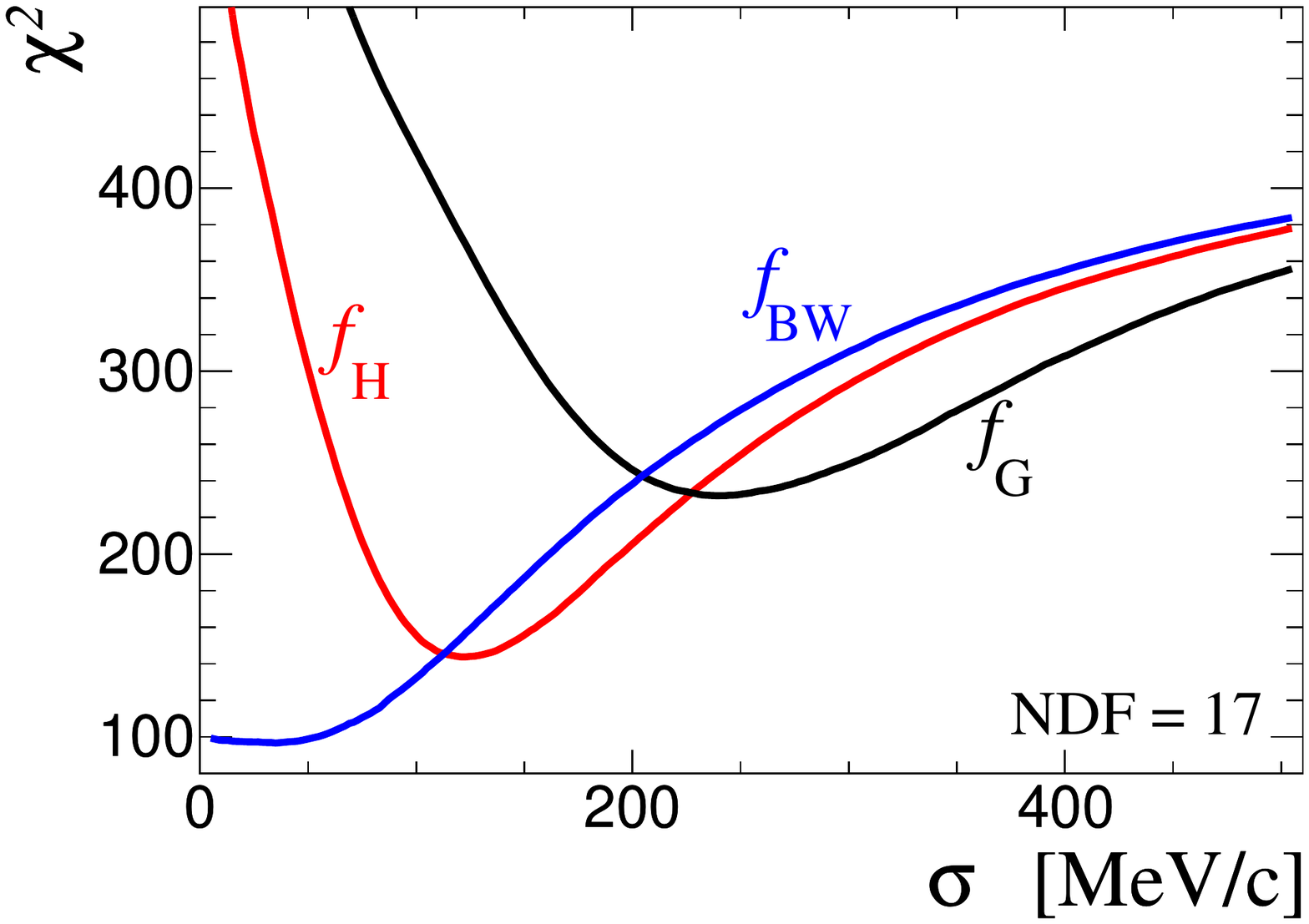}
  \end{center}
  \caption{\label{fig:Chi2}
    The deuteron breakup rate fit $\chi^2$ dependence on the nucleon momentum distribution function width $\sigma$.
  }
\end{figure}

The breakup rates in all 18 bins used (see Fig.\,\ref{fig:Rate}) were approximated following Eqs.\,(\ref{eq:omega}) and (\ref{eq:phase}) with $\Delta^d_\text{thr}\!=\!2.2\,\text{MeV}$. A function $\tilde{f}(-2m_pT_R,\Delta)$ [see Eq.\.(\ref{eq:dNdD})] was expressed via $f_\text{H}$, $f_\text{G}$, or $f_\text{BW}$, defined in the Appendix\,\ref{sec:func}. The fit $\chi^2$ dependence on the nucleon momentum distribution width parameter $\sigma$ is shown in Fig.\,\ref{fig:Chi2}.

Technically, the breakup fraction fit compares the isolated breakup event rate with the result of a two-step simulation. First, the elastic event is generated, i.e., the values of recoil proton energy $T_R$ (uniformly distributed) and $z$ coordinate in the Si detector (smeared in accordance with $\sigma_\text{jet}$) are assigned. Second, the mass excess $\Delta$ is calculated in accordance with the $\tilde{f}(T_R,\Delta)$ distribution (\ref{eq:dNdD}) and the $z$ coordinate is corrected following Eq.\,(\ref{eq:tanTheta}). The simulation normalization,
\begin{equation}
  \text{norm} = n_\text{el}(E_\text{beam})\times n_d(\tilde{f})
  \label{eq:norm}
\end{equation}
includes two factors, {\em(i)} the data sample (identified by the deuteron beam energy) dependent $n_\text{el}$, which is proportional to the elastic event statistics; and {\em(ii)} the nucleon momentum distribution function dependent $n_d\!\propto\!\tilde{f}(T_R,\Delta)d\Phi_2(T_R,\Delta)$, which is the same for all data samples.

The fit results are summarized in Table\,\ref{tab:dAu}. Four functions $\tilde{f}(T_R,\Delta)$ were probed. The first one is $f_\text{H}$ with fixed value of $\sigma\!=\!30\,\text{MeV/c}$\,\cite{Shindin:2017mcs}. In the other three, $f_\text{H}$, $f_\text{G}$, and $f_\text{BW}$ with $\sigma$ being a free parameter in the fit, were used.

\begin{table*}[t]
  \begin{center}
    \begin{tabular}{l|r r| c c | c c}
      $f(p_x,\sigma)$~~ &
      \multicolumn{1}{c}{~$\sigma\,\text{[MeV/c]}$~} &
      \multicolumn{1}{c|}{$\chi^2$} &
      $\omega_0$~[\%] & $\omega_1T_c$~[\%]~ & $\tilde{\omega}_0$~[\%] & $\tilde{\omega}_1T_c$~[\%] 
      \\[\smallskipamount]
      \hline
      $f_\text{H} $ &  30~~~~ &~~403.1~~~& $-0.92\phantom{-}$ & $0.76$ & $-0.10\phantom{-}$ & $0.08$ \\
      $f_\text{H} $ & 122~~~~ &  143.6~~~& $ 0.74$ & $0.20$ & $ 0.21$ & $0.05$ \\
      $f_\text{G} $ & 240~~~~ &  231.9~~~& $ 0.74$ & $0.15$ & $ 0.24$ & $0.05$ \\
      $f_\text{BW}$ &  35~~~~ &   96.6~~~& $ 0.37$ & $0.38$ & $ 0.10$ & $0.07$ \\[\smallskipamount]
      \hline
      \multicolumn{3}{r|}{$1/\chi^2$ weighted average~~~}& $~0.41\pm0.28~$ & $~0.33\pm0.10~$ & $~0.13\pm0.06~$ & $~0.06\pm0.01~$ 
    \end{tabular}
  \end{center}
  \caption{\label{tab:omega}
    An evaluation of the ${}^3\text{He}$ breakup fraction functions (\ref{eq:omega-nf}) and (\ref{eq:omega-sf}) based on the deuteron beam data fit within the recoil proton energy range of $1\!<\!T_R\!<\!7\,\text{MeV}$. $T_c$ is defined in Eq.\,(\ref{eq:Tc}). The specified errors should be attributed only to the result dependence on the function $\tilde{f}(T_R,\Delta)$ used in the fit.
  }
\end{table*}

Factor $n_d$ determined in the fit can be used to evaluate the breakup fraction
  $f_d\!=\!\langle\sigma_\text{qel}^{pd}/\sigma_\text{el}^{pd}\rangle$
for the 10--30\,GeV/nucleon deuteron beam and, then, to extrapolate it to the value of
$f_h\!=\!\langle\sigma_\text{qel}^{ph}/\sigma_\text{el}^{ph}\rangle$
in the ${}^3\text{He}$ beam measurements. Admitting that the result significantly depends on the model used, I calculated the average values, using weights inversely proportional to the values of $\chi^2$.  

Not-so-good ratios of $\chi^2/\text{NDF}$ may be related to {\em(i)} recoil proton shadowing in HJET, {\em(ii)} non-Gaussian tail in the jet density profile, {\em(iii)} the breakup rate dependence on the beam energy, and {\em(iv)}  possible dependence of $\psi(T_R,\Delta)$ on $T_R$ and $\Delta$. Potentially, {\em(i)} and {\em(ii)} can result in a significant bias in the evaluation of $n_d$. To check such a possibility, the same calculations were carried out using the accompanying Au beam events in the upper left {\em yellow} detector. A normalization factor $n_\text{Au}$ found is consistent with zero within statistical uncertainty of the measurements. This is an anticipated result due to much stronger phase space suppression for Au. Thus, one finds the experimental points drawn in Fig.\,\ref{fig:Rate} in {\em qualitative} agreement with the model used.  Since systematic uncertainties, mentioned above, are mostly the same for deuteron and gold beams, one can conclude that the breakup events were unambiguously isolated in the deuteron data.

\begin{figure}[t]
  \begin{center}
    \includegraphics[width=0.86\columnwidth]{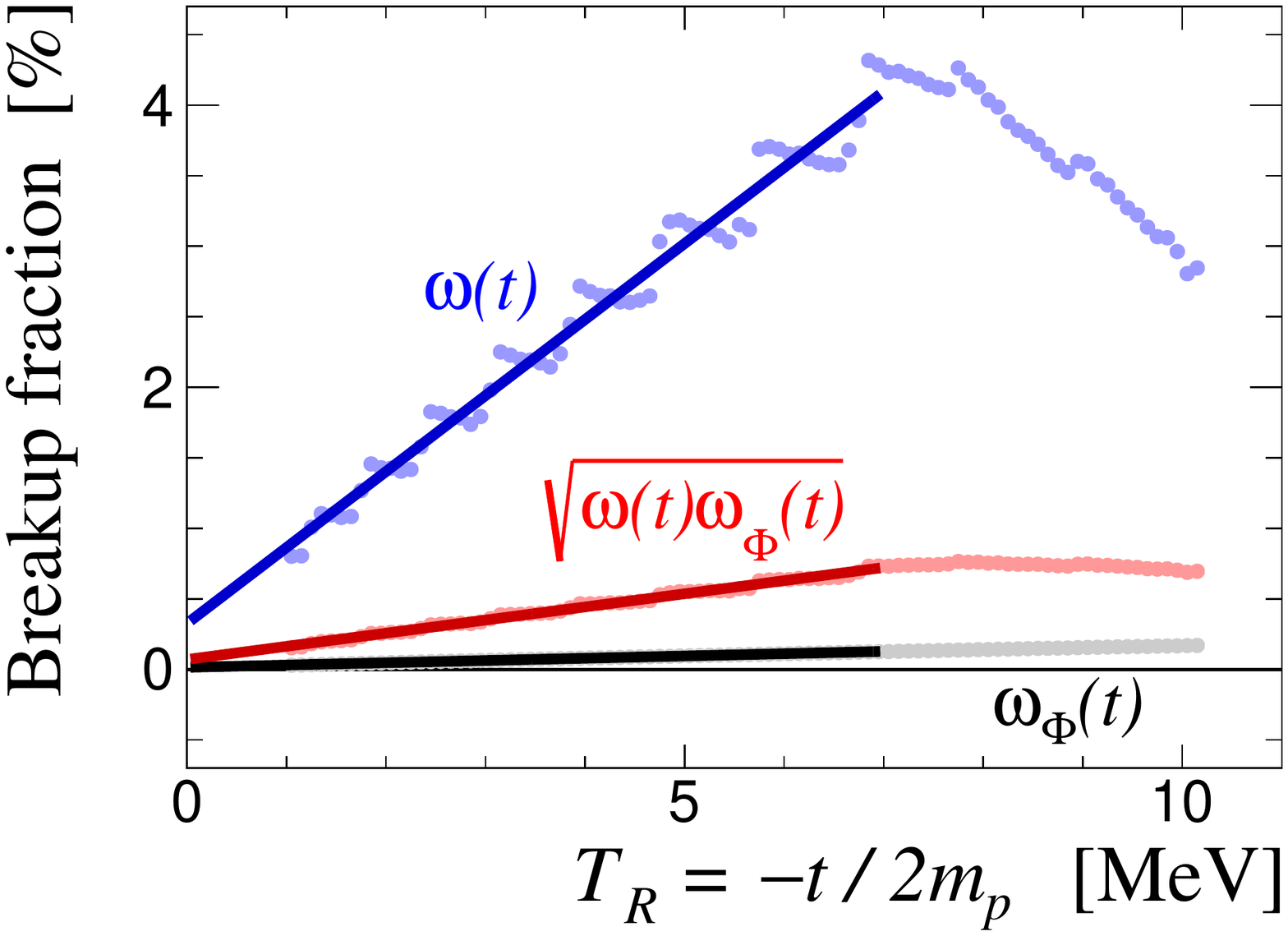}
  \end{center}
  \caption{\label{fig:omega}
  Expected ${}^3\text{He}$ breakup fraction dependence on $T_R$ derived from the deuteron beam data analysis. For the calculations, function $f_\text{BW}$ with $\sigma\!=\!35\,\text{MeV}$ was used. Nonsmooth dependence of the calculated points on $T_R$ reflects the discrete changes in the event selection efficiency due to the actual 3.75\,mm Si strip width. In the calculations, 100\,GeV/nucleon ${}^3\text{He}$ beam energy was assumed.
  }
\end{figure}

It was found (see Fig.\,\ref{fig:omega}) that the detected  breakup fraction linearly depends on $T_R$ (if $T_R\!<\!7\,\text{MeV}$). For $T_R\!>\!7\,\text{MeV}$, the linearity is broken due to finite size of the detector. Extrapolation of the low $T_R$ parametrization for the ${}^3\text{He}$ beam is given in Table\,\ref{tab:omega}. 

Interpreting the results obtained, one should note that
{\em(i)} the data analysis was based on an oversimplified and unjustified model;
{\em(ii)} only a statistically small part of the full $dN_\text{qel}/d\Delta$ distribution was available for the analysis;
{\em(iii)} the breakup fraction evaluated in the deuteron beam measurements was extrapolated to the helion beam.
Therefore, a verification of the estimate is critically important.

Breakup fractions $f_d$ and $f_h$ in Table\,\ref{tab:dAu} were calculated using typical HJET event selection cuts. For comparison with other estimates of breakup rates, the value of $\omega(T_R\!=\!3.5\,\text{MeV})$ was evaluated disregarding event selection cuts for $d$, ${}^3\rm{He}$, and ${}^6\rm{Li}$. Results were $5.0\!\pm\!1.5\,\%$, $2.7\!\pm\!0.4\,\%$, and $7.0\pm1.7\,\%$, respectively, The specified errors correspond to variation of $\omega$ depending on the choice of function $\tilde{f}(T_R,\Delta)$. Only phase space factor (\ref{eq:phase}) corrections were used to extrapolate the deuteron value to ${}^3\rm{He}$ and ${}^6\rm{Li}$. For ${}^6\text{Li}$ the following two-body breakups were considered: 
\begin{equation}
  {}^6\rm{Li} \to \left\{
  \begin{aligned}
    &{}^5\rm{Li}+n - 6.1\,\text{MeV},\\
    &{}^5\rm{He}+p - 4.8\,\text{MeV},\\
    &{}^4\rm{He}+d - 1.6\,\text{MeV},\\
    &{}^3\rm{H~}+h - 17.\,\text{MeV}. 
  \end{aligned}
  \right..
\end{equation}

Comparing the obtained result for ${}^3\rm{Li}$, $7.0\pm1.7\%$  with a value of 6.4\% evaluated from Fig.\,1 of Ref.\,\cite{Glauber:1970jm}, one finds excellent agreement within the specified uncertainty of about 25\%.

So, regardless of important experimental and theoretical uncertainties in the evaluation of the ${}^3\rm{He}$ breakup rate, there is no indication that the breakup fraction found may be significantly (e.g., by a factor of 10) underestimated. As will be shown below, even an order-of-magnitude underestimate of the breakup fraction does not discard the conclusion that the breakup effect is negligible for the helion beam polarization measurement based on Eq.\,(\ref{eq:PolBeamR}).

\section{
The breakup corrections to the measured $\boldsymbol{}^3\text{He}$ beam polarization
}

\subsection{Single spin $\boldsymbol{h^\uparrow\!\!\!p}$ analyzing power}

Polarization effects in the forward scattering of a polarized spin-1/2 beam particle off a, generally non-identical, unpolarized target is dominantly described by two helicity amplitudes\,\cite{Buttimore:1998rj}: nonflip $\phi_+(s,t)$ and spin-flip $\phi_5(s,t)$. In the CNI region, both hadronic and electromagnetic components are essential, $\phi\!=\!\phi^\text{had}\!+\!\phi^\text{em}e^{i\delta_C}$, where QED calculated $\phi^\text{em}$ is considered to be real and $\delta_C\!\approx\!0.02$ is a Coulomb phase\,\cite{Cahn:1982nr,Kopeliovich:2000ez,Poblaguev:2021xkd}. In the considered approach,
\begin{equation}
  A_\text{N}(s,t) = {-2\text{Im}\left(\phi_5^*\phi_+\right)}/{|\phi_+|^2}.
  \label{eq:AN_ampl}
\end{equation}

Following Ref.\,\cite{Poblaguev:2022yzw} and omitting some corrections which are inessential in the context of this paper, but may be needed for a precision measurement\,\cite{Poblaguev:2019vho}, the ${h^\uparrow{p}}$ analyzing power can be parametrized\,\cite{Poblaguev:2022yzw} as
\begin{eqnarray}
  A_\text{N}(T_R) &=& A_\text{N}^\text{nf}(T_R)\times
  \left[\varkappa_h\!-\!0.54\,I_5\!-\!0.54\,R_5T_R/T_c\right],\qquad
  \label{eq:AN} \\
  A_\text{N}^\text{nf}(T_R) &=& \sqrt{\frac{2T_R}{m_p}}\,\frac{1}{F_\text{cs}(T_R)},\\
  F_\text{cs}(T_R) &=& {T_c/T_R \!+\!\beta_0\!+\!1\!+\!\beta_1\!T_R/T_c+\!\beta_2(T_R/T_c)^2},
  \label{eq:AN_}
\end{eqnarray}
where $\varkappa_hA_\text{N}^\text{nf}$ is the analyzing power calculated assuming $r_5\!=\!0$, $F_\text{cs}(-t/2m_p)\!\propto\!d\sigma^{hP}/dt\!\times\!t/t_c$ is defined by the ${hp}$ differential cross section parametrization\,\cite{Poblaguev:2022yzw}, and
\begin{align}
  \varkappa_h &= \mu_h/Z_h-m_p/m_h~  & &\hspace{-4em}= -1.398, \label{eq:kappa}  \\
  T_c\,       &=-4\pi\alpha Z_h/m_p\sigma^{hp}_\text{tot}~\,  & &\hspace{-4em}\approx 0.74\,\text{MeV},
  \label{eq:Tc} \\
  \beta_0\:   &= -2\,(\rho_{hp}+ \delta_C^{hp}) &  &\hspace{-4em}\approx 0.1, \\
  \beta_1\:   &= \beta_2                  &  &\hspace{-4em}\approx 0,
\end{align}
$\sigma^{hp}_\text{tot}$ is the total $hp$ cross section, $\rho_{hp}$ is the forward real to imaginary ratio, and $r_5\!=\!R_5\!+\!iI_5$ is the hadronic spin-flip amplitude parameter\,\cite{Buttimore:1998rj} measured in ${pp}$ scattering\,\cite{Poblaguev:2019saw}. The numerical estimates are given for $E_\text{beam}\!=\!100\,\text{GeV/nucleon}$. In Eq.\,(\ref{eq:AN}), parameters $\varkappa_h$, $I_5$, and $R_5$ can be attributed to Eq.\,(\ref{eq:AN_ampl}) terms $\phi_5^\text{em}\,\text{Im}\phi_+^\text{had}$, $\phi_+^\text{em}\,\text{Im}\phi_5^\text{had}$, and $\text{Re}\phi_5^\text{had}\,\text{Im}\phi_+^\text{had}$, respectively.

For the unpolarized hydrogen target, the helium beam polarization (as a function of the recoil proton energy $T_R$) can be measured if the analyzing power is known,
\begin{eqnarray}
  P_\text{meas}(T_R) &=& a_\text{beam}/A_\text{N}(T_R) \nonumber \\
  &=& P_\text{beam}\times\left(1+\delta{\xi_0}+\delta{\xi_1}T_R/T_c\right),
  \label{eq:Pmeas}
\end{eqnarray}
where $\delta{\xi_{0,1}}$ are systematic errors due to possible uncertainties in $A_\text{N}(T_R)$\,\cite{Poblaguev:2022yzw}. Since $\delta{\xi_1}$ can be determined directly in a linear fit of Eq.\,(\ref{eq:Pmeas}), $A_\text{N}$ related contribution to the measured beam polarization systematic error is defined by $\delta{\xi_0}$ only
\begin{equation}
  \delta P_\text{beam}/P_\text{beam} = \delta{\xi_0}.
\end{equation}

\subsection{
  The $\boldsymbol{{}^3}\text{He}$ breakup corrections to $\boldsymbol{|\phi_+|^2}$
}

According to estimates done in Ref.\,\cite{Poblaguev:2022yzw}, possible uncertainties in parametrization of $F_\text{cs}(T_R)$ will effectively result in
\begin{equation}
  \delta{\xi_0} = 0.49\,\delta{\beta_0} + 0.58\,\delta{\beta_1}
  - 0.82\,\delta{\beta_2}. 
\end{equation}

As evaluated above the effective helion breakup correction to the ${hp}$ elastic cross section\,(\ref{eq:omega}) leads to
\begin{equation}
  \delta_{|\phi_+|^2}^\text{qel} P/P= 0.58\,\omega_0
  - 0.82\,\omega_1T_c \approx -0.03\pm0.18,\%
  \label{eq:csCorr}
\end{equation}
where the specified error was derived from uncertainties in values of  $\omega_{0,1}$ in Table\,\ref{tab:omega}. As written, the result obtained is consistent with the EIC requirement (\ref{eq:systEIC}).
However, {\em(i)} the model-dependent uncertainties in Eq.\,(\ref{eq:csCorr}) are not reliably determined; {\em(ii)} the $T_R$ range used in the deuteron beam data analysis did not allow evaluation of the breakup correction to $\beta_0$, e.g., in pure electromagnetic scattering; and {\em(iii)} the considered uncertainties in values of $\beta_1$ and $\beta_2$ may affect the accuracy of an experimental determination of $\beta_0$ in the $d\sigma/dt$ data fit\,\cite{Poblaguev:2022yzw}.

Thus, for the ${}^3\text{He}$ beam polarization measurement based on Eq.\,(\ref{eq:Pmeas}), one cannot exclude large systematic error
\begin{equation}
  \delta_{|\phi_+|^2}^\text{qel} P/P \gtrsim 1\%
  \label{eq:csCorrRev}
\end{equation}
due to the breakup corrections to the ${hp}$ cross section.

\subsection{
   The breakup correction to the spin-flip $\boldsymbol{\phi_5^\text{em}\text{Im}\,\phi_+^\text{had}}$ interference term
}
  
The effective (i.e., including inelastic component) analyzing power $A_\text{N}^\text{eff}(t)$ can be derived from Eqs.\,(\ref{eq:AN}) and (\ref{eq:AN_}) by adding the effective breakup amplitudes. 

Interference of the spin-flip electromagnetic and non-flip hadronic amplitudes,
which gives dominant contribution to the analyzing power, will be modified as
\begin{equation}
  \varkappa_h\quad\to\quad \varkappa_h\times\left[1 + \tilde{\omega}_\text{nf}(t)\right],
\end{equation}
where, similarly to Eqs.\,(\ref{eq:omega}) and (\ref{eq:phase}),
\begin{align}
  &\!\!\tilde{\omega}_\text{nf}(t) = \frac{\sqrt{2m_pm_d}}{4\pi m_h}\times \nonumber \\
  &\!\int_{\Delta^h_\text{thr}}^{\infty}{d\Delta\,\text{Im}\big[(1\!+\!\delta_\varkappa)(i\!+\!\rho_{hp})\psi\big]\,\tilde{f}(t,\Delta)\sqrt{ \frac{\Delta\!-\!\Delta^h_\text{thr}}{m_h} } }.
  \label{eq:I*+}
\end{align}
Here, $\psi(t,\Delta)$ is defined in Eq.\,(\ref{eq:phi_brk}) and $\varkappa_h\delta_\varkappa(t,\Delta)$ is a correction to the electromagnetic spin-flip amplitude due to the ${}^3\text{He}$ breakup. Using Eq.\,(\ref{eq:kappa}), one can evaluate
\begin{equation}
  \delta_\varkappa \approx \Delta/2m_h = {\cal O}(10^{-3}).
\end{equation}
Neglecting second order corrections and assuming that $\psi(t,\Delta)$ may have only a week dependence on $\Delta$ and $\text{Im}\left[(i+\rho)\psi\right]\!\approx\!|(i+\rho)\psi|$ (which maximize $\tilde{\omega}_\text{nf}$), one finds that breakup correction $\tilde{\omega}_\text{nf}$  to the ${\phi_5^\text{em}\text{Im}\,\phi_+^\text{had}}$ interference term,
\begin{equation}
  \left|\tilde{\omega}_\text{nf}(t)\right| \le |\tilde{\omega}(t)|
  \label{eq:I5*}
\end{equation}
can be limited by function defined in Eq.\,(\ref{eq:omega-sf}).

Subsequent systematic error in the measured beam polarization is
\begin{equation}
  |\delta_{\phi_5^\text{em}\phi_+^\text{had}}^\text{qel}P/P| \le \tilde{\omega}_0 \approx 0.13\pm0.06\,\%,
\end{equation}
which may be considered as small.

\subsection{The polarized target measurements}

To evaluate the breakup-related uncertainty $\delta^\text{qel}\xi_0$ in the helion beam polarization measurement at HJET, one can, using corrections discussed above, rewrite Eq.\,(\ref{eq:PolBeamR}) as
\begin{align}
  \!\!\!&P_\text{beam}^h = P_\text{jet}\,\frac{a_\text{beam}(T_R)}{a_\text{jet}(T_R)}
  \nonumber\\
  &\times\frac%
      {\varkappa_p[1\!+\!\tilde{\omega}_\text{nf}(T_R)]-2I_5[1\!+\!\tilde{\omega}_I^p(T_R)]-2R_5T_R/T_c}%
      {\varkappa_h[1\!+\!\tilde{\omega}_\text{nf}(T_R)]-0.54I_5[1\!+\!\tilde{\omega}_I^h(T_R)]-0.54R_5T_R/T_c},
      \label{eq:P_h/p}
\end{align}
where $\varkappa_p\!=\!1.793$ is the anomalous magnetic moment of a proton and $\tilde{\omega}_I^{p,h}$ are effective breakup corrections to the hadronic spin-flip $p^\uparrow{h}$ and $h^\uparrow{p}$ amplitudes, respectively. Consequently,
\begin{equation}
  \delta^\text{qel}\xi_0 = \left[%
  \frac{2I_5}{\varkappa_p}(\tilde{\omega}_\text{nf}\!-\!\tilde{\omega}_I^p) - 
  \frac{0.54I_5}{\varkappa_h}(\tilde{\omega}_\text{nf}\!-\!\tilde{\omega}_I^h)
  \right]_{T_R\to0}.
\end{equation}
Hypothesizing that, similarly to the elastic scattering\,\cite{Kopeliovich:2000kz}, the ratio of the spin-flip to the nonflip parts of a breakup proton-nucleus amplitude (e.g., for $h^\uparrow p\to pd\,p$) is the same as for elastic proton-proton amplitude, one can expect $\tilde{\omega}_I^{p,h}(T_R)\!=\!\tilde{\omega}_\text{nf}(T_R)$ and, consequently,  $\delta^\text{qel}\xi_0\!=\!0$.

Here, to test the robustness, a more conservative estimate $|\tilde{\omega}_\text{nf}(T_R)\!-\!\tilde{\omega}_I^{p,h}(T_R)|\!\lesssim\!\tilde{\omega}_\text{nf}^\text{max}\!=\!0.05$ was considered. The upper limit $\tilde{\omega}_\text{nf}^\text{max}$ used is about a factor of 7 larger than that displayed in Fig.\,\ref{fig:omega} for $\tilde{\omega}(T_R)$. Such a value of $\tilde{\omega}_\text{nf}^\text{max}$ , if attributed to a possible error in the evaluation of the breakup rate, suggests that the breakup fraction was underestimated by a factor of about 50. Considering $|r_5|\!=\!r_5^\text{max}\!\approx\!0.02$, one finds $|\delta^\text{qel}\xi_0|\!\lesssim\!r_5^\text{max}\tilde{\omega}_\text{nf}^\text{max}\!\approx\!10^{-3}$. However, it was implicitly assumed that $|\tilde{\omega}_\text{nf}(T_R)|\!=\!\tilde{\omega}_\text{nf}^\text{max}$ in this estimate. In a more reasonable approximation, $|\tilde{\omega}_\text{nf}(T_R)|\!=\!\tilde{\omega}_\text{nf}^\text{max}T_R/T_R^\text{max}$, where $T_R^\text{max}\!=\!10\,\text{MeV}$, one comes to $|\delta^\text{qel}\xi_0|\!\to\!0$ and $|\delta^\text{qel}\xi_1|\!\lesssim\!r_5^\text{max}\tilde{\omega}_\text{nf}T_c/T_R^\text{max}$.

A larger value of  $\delta^\text{qel}\xi_1$ follows from the breakup corrections to $R_5$
\begin{equation}
 \delta^\text{qel}\xi_1\frac{T_R}{T_c} \approx
 \left(-\frac{2}{\varkappa_p}+\frac{0.54}{\varkappa_h}\right)
 \Big\langle\omega(T_R)\!-\!\tilde{\omega}_\text{nf}(T_R)\Big\rangle
   \,R_5\frac{T_R}{T_c},
\end{equation}
where $\omega(T_R)$ is shown in Fig.\,\ref{fig:omega} and $\omega(T_R)\!-\!\tilde{\omega}_\text{nf}(T_R)$ is averaged over $T_R$ range considered.

Thus, only a small correction to the measured polarization $P_\text{meas}(T_R)$ [Eq.\,(\ref{eq:Pmeas})] due to the ${}^3\text{He}$ breakup was found
\begin{equation}
  |\delta^\text{qel}\xi_0| \ll      0.1\%, \qquad
  |\delta^\text{qel}\xi_1| = {\cal O}(0.1\%).
\end{equation}
  
\section{Summary}

Investigating a possible effect of the ${}^3\text{He}$ breakup in the helion beam polarization measurement at EIC, I came to a counterintuitive conclusion that the effect cancels and, thus, can be disregarded in the HJET measurements based on the concurrent determination of the beam and target (jet) spin-correlated asymmetries as explained in Eq.\,(\ref{eq:PolBeamR}).

Since only low energy, $T_R\!<\!10\,\text{MeV}$, recoil protons can be detected at HJET, the breakup events are kinematically indistinguishable from the elastic ones for the high energy, $\sim\!100\,\text{GeV/nucleon}$, helion beam. However, for the detected events,  the breakup rate is expected to be only a few percent of the elastic one. Also the effective breakup corrections to $A_\text{N}^{hp}(t)$ and $A_\text{N}^{ph}(t)$ cancel,  to about 0.1\% level, in the analyzing power ratio (\ref{eq:PolBeamR}). Thus, anticipated systematic error in $P_\text{beam}^h$ due to the ${}^3\text{He}$ breakup can be neglected compared the EIC requirement (\ref{eq:systEIC}). 

Summarizing, to determine the helion beam absolute polarization using HJET, one should measure the beam $a_\text{beam}^h$ and jet $a_\text{jet}^p$ spin-correlated asymmetries and then, calculate the beam polarization as
\begin{align}
&P_\text{beam}^h = P_\text{jet}\,{a_\text{beam}^h}(T_R)/{a_\text{jet}^p(T_R)} \nonumber \\
      &\qquad\times \frac%
      {\varkappa_p-m_p^2/m_hE_\text{beam}-2I_5-2R_5T_R/T_c}%
      {\varkappa_h-m_h/E_\text{beam}-0.54I_5-0.54R_5T_R/T_c}.
      \label{eq:P_h/p}
\end{align}
  A detailed explanation of Eq.\,(\ref{eq:P_h/p}), including the beam energy per nucleon, $E_\text{beam}$, dependent corrections, not discussed  here, is given in Ref.\,\cite{Poblaguev:2022yzw}.

The conclusion was found to be stable against possible model-dependent uncertainties in the evaluation of the breakup rate. Nonetheless, it should be underlined that the ${}^3\text{He}$ breakup rate was approximated by a simplified extrapolation of the partially measured deuteron beam $dN/d\Delta$ distribution to the helion rate integrated over the full range in $\Delta$.
  
The estimate of the ${}^3\text{He}$ breakup rate can be significantly improved in a designated $\sim\!1$ day (unpolarized) helion beam measurements at HJET (before RHIC will be closed for the EIC construction). To maximize the effect, the beam energy should be 10.6\,GeV/nucleon (the ${}^3\text{He}$ injection energy to RHIC). To minimize systematic errors in this study, only one RHIC beam should be loaded and the HJET holding field magnet should be turned off. Additional similar measurements for the helion beam energies $\approx\!5$ and $\approx\!20\,\text{GeV/nucleon}$ may allow one to determine the full $dN/d\Delta$ distribution. The method suggested has already been tested using an Au beam\,\cite{Au_Breakup}.

\acknowledgements{
  A significant contribution to the experimental part of this work, deuteron beam measurements, was given by Anatoli Zelenski and Grigor Atoian.  The author would like to thank Nigel Buttimore and Anatoli Zelenski for useful discussions and Boris Kopeliovich for valuable comments, and acknowledges support from the Office of Nuclear Physics in the Office of Science of the U.S. Department of Energy. This work is authored by employee of Brookhaven Science Associates, LLC under Contract No. DE-SC0012704 with the U.S. Department of Energy.
}

\appendix
\section{\label{sec:func}
  Nucleon momentum distribution function
}

For an evaluation of the nucleon momentum distribution (\ref{eq:dNdD}) in a nuclear, one can utilize a Gaussian function
\begin{equation}
  f_\text{G}(p_x,\sigma) = C\times\displaystyle
  \frac{\exp{(-p_x^2/2\sigma^2)}}{\sqrt{2\pi}\sigma}
  \label{eq:fG}
\end{equation}
or a Breit--Wigner function 
\begin{equation}
  f_\text{BW}(p_x,\sigma) = C\times
  \frac{\pi^{-1}\,\sqrt{2}\sigma}{p_x^2+2\sigma^2}.
  \label{eq:fBW}
\end{equation}
\begin{figure}[t]
  \begin{center}
    \includegraphics[width=0.85\columnwidth]{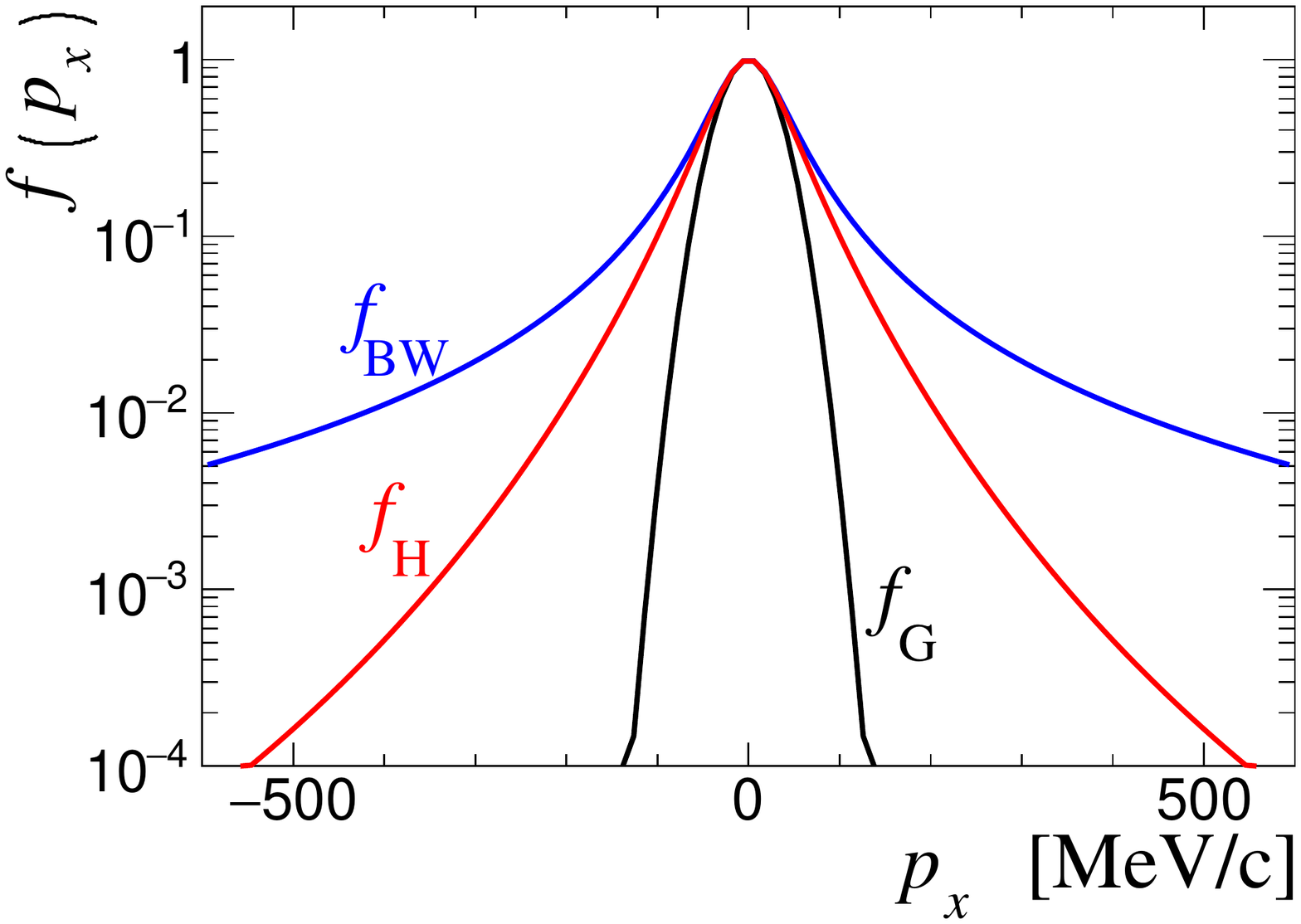}
  \end{center}
  \caption{\label{fig:func}
    Functions $f_\text{G}(p_x,\sigma)$, $f_\text{H}(p_x,\sigma)$, and $f_\text{BW}(p_x,\sigma)$
  normalized to unity at maximum. $\sigma\!=\!\sigma_d\!=\!30\,\text{MeV/c}$.
  }
\end{figure}

For a deuteron, such a distribution can be derived using the Hulth\'{e}n wave function parametrization\,\cite{Hulthen:1957}
\begin{equation}
  \Psi(r) \propto \left(e^{-\alpha r}-e^{-\beta r}\right)/r.
  \label{eq:Psi}
\end{equation}
After Fourier transformation and integration, one finds
\begin{align}
  f_\text{H}&(p_x,\alpha,\beta) =
  \frac{\alpha\beta(\alpha+\beta)}{\pi(\beta\!-\!\alpha)^2} \nonumber \\
  &\;\times\left[\frac{1}{\alpha^2\!+\!p_x^2}\!+\!\frac{1}{\beta^2\!+\!p_x^2}\!-\!%
  \frac{2}{\beta^2\!-\!\alpha^2}\,\ln{\frac{\beta^2\!+\!p_x^2}{\alpha^2\!+\!p_x^2}}\right].
\end{align}
Following Ref.\,\cite{Shindin:2017mcs}, one should assume
\begin{equation}
  \alpha=45.7\,\text{MeV/c}\text{ and }\beta=260\,\text{MeV/c},
  \label{eq:alpha,beta}
\end{equation}
which are consistent with results obtained in the experimental study of the deuteron breakup in Dubna\,\cite{Atanasov:1994}.

Function $f_\text{H}$ can be rewritten in a form similar to (\ref{eq:fG}) and (\ref{eq:fBW}),  
\begin{equation}
  f_\text{H}(p_x,\sigma)=C\times f_\text{H}(p_x,\alpha\sigma/\sigma_d,\beta\sigma/\sigma_d),
  \label{eq:fH}
\end{equation}
where
\begin{eqnarray}
  \sigma_d&=& \frac{\alpha^{-2}+\beta^{-2}-2\left(\beta^2\!-\!\alpha^2\right)^{-1}\,\ln{\left(\beta^2/\alpha^2\right)}}%
        {\left(\alpha^{-2}-\beta^{-2}\right)^2} \nonumber \\
        &\approx& 30\,\text{MeV/c}.
\end{eqnarray}

Functions $f_\text{G}(p_x,\sigma)$, $f_\text{H}(p_x,\sigma)$, and $f_\text{G}(p_x,\sigma)$ are depicted in Fig.\,\ref{fig:func}. All functions have the same behavior around the maximum at $p_x\!=\!0$
\begin{equation}
  f(p_x,\sigma)\approx f(0,\sigma)\times\left(1-p_x^2/2\sigma^2\right).
\end{equation}
Since the functions are unity integral normalized (if $C\!=\!1$) and the distribution widths are proportional to $\sigma$ they can be used to approximate the Dirac $\delta$ function
\begin{equation}
  \lim_{\sigma\to0} f(p_x,\sigma) = C\,\delta(p_x).
\end{equation}

In Hulth\'{e}n approximation (\ref{eq:Psi}), parameter $\alpha$ is fixed by the deuteron binding energy $B\!=\!2.22\,\text{MeV}$\,\cite{Hulthen:1957},
\begin{equation}
  \alpha = \sqrt{Bm_p}=45.7\,\text{MeV}.
\end{equation}
Therefore, $f_\text{H}(p_x,\sigma)$ is not supposed to be used to fit nucleon momentum distribution in deuteron. In this paper, $f_\text{H}$ is considered as a generic function that is similar to $f_\text{G}$ and $f_\text{BW}$ around the maximum but has an intermediate width compared to Gaussian and Breit-Wigner distributions.


\begin{thebibliography}{25}%
\makeatletter
\providecommand \@ifxundefined [1]{%
 \@ifx{#1\undefined}
}%
\providecommand \@ifnum [1]{%
 \ifnum #1\expandafter \@firstoftwo
 \else \expandafter \@secondoftwo
 \fi
}%
\providecommand \@ifx [1]{%
 \ifx #1\expandafter \@firstoftwo
 \else \expandafter \@secondoftwo
 \fi
}%
\providecommand \natexlab [1]{#1}%
\providecommand \enquote  [1]{``#1''}%
\providecommand \bibnamefont  [1]{#1}%
\providecommand \bibfnamefont [1]{#1}%
\providecommand \citenamefont [1]{#1}%
\providecommand \href@noop [0]{\@secondoftwo}%
\providecommand \href [0]{\begingroup \@sanitize@url \@href}%
\providecommand \@href[1]{\@@startlink{#1}\@@href}%
\providecommand \@@href[1]{\endgroup#1\@@endlink}%
\providecommand \@sanitize@url [0]{\catcode `\\12\catcode `\$12\catcode
  `\&12\catcode `\#12\catcode `\^12\catcode `\_12\catcode `\%12\relax}%
\providecommand \@@startlink[1]{}%
\providecommand \@@endlink[0]{}%
\providecommand \url  [0]{\begingroup\@sanitize@url \@url }%
\providecommand \@url [1]{\endgroup\@href {#1}{\urlprefix }}%
\providecommand \urlprefix  [0]{URL }%
\providecommand \Eprint [0]{\href }%
\providecommand \doibase [0]{https://doi.org/}%
\providecommand \selectlanguage [0]{\@gobble}%
\providecommand \bibinfo  [0]{\@secondoftwo}%
\providecommand \bibfield  [0]{\@secondoftwo}%
\providecommand \translation [1]{[#1]}%
\providecommand \BibitemOpen [0]{}%
\providecommand \bibitemStop [0]{}%
\providecommand \bibitemNoStop [0]{.\EOS\space}%
\providecommand \EOS [0]{\spacefactor3000\relax}%
\providecommand \BibitemShut  [1]{\csname bibitem#1\endcsname}%
\let\auto@bib@innerbib\@empty
%</preamble>
\bibitem [{\citenamefont {Accardi}\ \emph {et~al.}(2016)\citenamefont
  {Accardi}, \citenamefont {Albacete}, \citenamefont {Anselmino} \emph
  {et~al.}}]{Accardi:2012qut}%
  \BibitemOpen
  \bibfield  {author} {\bibinfo {author} {\bibfnamefont {A.}~\bibnamefont
  {Accardi}}, \bibinfo {author} {\bibfnamefont {L.}~\bibnamefont {Albacete},
  \bibfnamefont {J.}}, \bibinfo {author} {\bibfnamefont {M.}~\bibnamefont
  {Anselmino}}, \emph {et~al.},\ }\bibfield  {title} {\bibinfo {title}
  {{Electron Ion Collider: The Next QCD Frontier}},\ }\href
  {https://doi.org/10.1140/epja/i2016-16268-9} {\bibfield  {journal} {\bibinfo
  {journal} {Eur. Phys. J. A}\ }\textbf {\bibinfo {volume} {52}},\ \bibinfo
  {pages} {268} (\bibinfo {year} {2016})},\ \Eprint
  {https://arxiv.org/abs/1212.1701} {arXiv:1212.1701 [nucl-ex]} \BibitemShut
  {NoStop}%
\bibitem [{\citenamefont {Abdul~Khalek}\ \emph {et~al.}(2022)\citenamefont
  {Abdul~Khalek} \emph {et~al.}}]{AbdulKhalek:2021gbh}%
  \BibitemOpen
  \bibfield  {author} {\bibinfo {author} {\bibfnamefont {R.}~\bibnamefont
  {Abdul~Khalek}} \emph {et~al.},\ }\bibfield  {title} {\bibinfo {title}
  {{Science Requirements and Detector Concepts for the Electron-Ion Collider}:
  {EIC Yellow Report}},\ }\href
  {https://doi.org/10.1016/j.nuclphysa.2022.122447} {\bibfield  {journal}
  {\bibinfo  {journal} {Nucl. Phys. A}\ }\textbf {\bibinfo {volume} {1026}},\
  \bibinfo {pages} {122447} (\bibinfo {year} {2022})},\ \Eprint
  {https://arxiv.org/abs/2103.05419} {arXiv:2103.05419 [physics.ins-det]}
  \BibitemShut {NoStop}%
\bibitem [{\citenamefont {Zelenski}\ \emph {et~al.}(2005)\citenamefont
  {Zelenski}, \citenamefont {Bravar}, \citenamefont {Graham}, \citenamefont
  {Haeberli}, \citenamefont {Kokhanovski}, \citenamefont {Makdisi},
  \citenamefont {Mahler}, \citenamefont {Nass}, \citenamefont {Ritter},
  \citenamefont {Wise},\ and\ \citenamefont {Zubets}}]{Zelenski:2005mz}%
  \BibitemOpen
  \bibfield  {author} {\bibinfo {author} {\bibfnamefont {A.}~\bibnamefont
  {Zelenski}}, \bibinfo {author} {\bibfnamefont {A.}~\bibnamefont {Bravar}},
  \bibinfo {author} {\bibfnamefont {D.}~\bibnamefont {Graham}}, \bibinfo
  {author} {\bibfnamefont {W.}~\bibnamefont {Haeberli}}, \bibinfo {author}
  {\bibfnamefont {S.}~\bibnamefont {Kokhanovski}}, \bibinfo {author}
  {\bibfnamefont {Y.}~\bibnamefont {Makdisi}}, \bibinfo {author} {\bibfnamefont
  {G.}~\bibnamefont {Mahler}}, \bibinfo {author} {\bibfnamefont
  {A.}~\bibnamefont {Nass}}, \bibinfo {author} {\bibfnamefont {J.}~\bibnamefont
  {Ritter}}, \bibinfo {author} {\bibfnamefont {T.}~\bibnamefont {Wise}},\ and\
  \bibinfo {author} {\bibfnamefont {V.}~\bibnamefont {Zubets}},\ }\bibfield
  {title} {\bibinfo {title} {{Absolute polarized H-jet polarimeter development,
  for RHIC}},\ }\href {https://doi.org/10.1016/j.nima.2004.08.080} {\bibfield
  {journal} {\bibinfo  {journal} {Nucl. Instrum. Meth. A}\ }\textbf {\bibinfo
  {volume} {536}},\ \bibinfo {pages} {248} (\bibinfo {year}
  {2005})}\BibitemShut {NoStop}%
\bibitem [{\citenamefont {Poblaguev}\ \emph {et~al.}(2020)\citenamefont
  {Poblaguev}, \citenamefont {Zelenski}, \citenamefont {Atoian}, \citenamefont
  {Makdisi},\ and\ \citenamefont {Ritter}}]{Poblaguev:2020qbw}%
  \BibitemOpen
  \bibfield  {author} {\bibinfo {author} {\bibfnamefont {A.~A.}\ \bibnamefont
  {Poblaguev}}, \bibinfo {author} {\bibfnamefont {A.}~\bibnamefont {Zelenski}},
  \bibinfo {author} {\bibfnamefont {G.}~\bibnamefont {Atoian}}, \bibinfo
  {author} {\bibfnamefont {Y.}~\bibnamefont {Makdisi}},\ and\ \bibinfo {author}
  {\bibfnamefont {J.}~\bibnamefont {Ritter}},\ }\bibfield  {title} {\bibinfo
  {title} {{Systematic error analysis in the absolute hydrogen gas jet
  polarimeter at RHIC}},\ }\href {https://doi.org/10.1016/j.nima.2020.164261}
  {\bibfield  {journal} {\bibinfo  {journal} {Nucl. Instrum. Meth. A}\ }\textbf
  {\bibinfo {volume} {976}},\ \bibinfo {pages} {164261} (\bibinfo {year}
  {2020})},\ \Eprint {https://arxiv.org/abs/2006.08393} {arXiv:2006.08393
  [physics.ins-det]} \BibitemShut {NoStop}%
\bibitem [{\citenamefont {Buttimore}\ \emph {et~al.}(1999)\citenamefont
  {Buttimore}, \citenamefont {Kopeliovich}, \citenamefont {Leader},
  \citenamefont {Soffer},\ and\ \citenamefont {Trueman}}]{Buttimore:1998rj}%
  \BibitemOpen
  \bibfield  {author} {\bibinfo {author} {\bibfnamefont {N.~H.}\ \bibnamefont
  {Buttimore}}, \bibinfo {author} {\bibfnamefont {B.~Z.}\ \bibnamefont
  {Kopeliovich}}, \bibinfo {author} {\bibfnamefont {E.}~\bibnamefont {Leader}},
  \bibinfo {author} {\bibfnamefont {J.}~\bibnamefont {Soffer}},\ and\ \bibinfo
  {author} {\bibfnamefont {T.~L.}\ \bibnamefont {Trueman}},\ }\bibfield
  {title} {\bibinfo {title} {{The spin dependence of high-energy proton
  scattering}},\ }\href {https://doi.org/10.1103/PhysRevD.59.114010} {\bibfield
   {journal} {\bibinfo  {journal} {Phys. Rev. D}\ }\textbf {\bibinfo {volume}
  {59}},\ \bibinfo {pages} {114010} (\bibinfo {year} {1999})},\ \Eprint
  {https://arxiv.org/abs/hep-ph/9901339} {arXiv:hep-ph/9901339} \BibitemShut
  {NoStop}%
\bibitem [{\citenamefont {Poblaguev}(2019)}]{Poblaguev:2019vho}%
  \BibitemOpen
  \bibfield  {author} {\bibinfo {author} {\bibfnamefont {A.~A.}\ \bibnamefont
  {Poblaguev}},\ }\bibfield  {title} {\bibinfo {title} {{Corrections to the
  Elastic Proton-Proton Analyzing Power Parametrization at High Energies}},\
  }\href {https://doi.org/10.1103/PhysRevD.100.116017} {\bibfield  {journal}
  {\bibinfo  {journal} {Phys. Rev. D}\ }\textbf {\bibinfo {volume} {100}},\
  \bibinfo {pages} {116017} (\bibinfo {year} {2019})},\ \Eprint
  {https://arxiv.org/abs/1910.02563} {arXiv:1910.02563 [hep-ph]} \BibitemShut
  {NoStop}%
\bibitem [{\citenamefont {Poblaguev}\ \emph {et~al.}(2019)\citenamefont
  {Poblaguev}, \citenamefont {Zelenski}, \citenamefont {Aschenauer},
  \citenamefont {Atoian}, \citenamefont {Eyser}, \citenamefont {Huang},
  \citenamefont {Makdisi}, \citenamefont {Schmidke}, \citenamefont {Alekseev},
  \citenamefont {Svirida},\ and\ \citenamefont
  {Buttimore}}]{Poblaguev:2019saw}%
  \BibitemOpen
  \bibfield  {author} {\bibinfo {author} {\bibfnamefont {A.~A.}\ \bibnamefont
  {Poblaguev}}, \bibinfo {author} {\bibfnamefont {A.}~\bibnamefont {Zelenski}},
  \bibinfo {author} {\bibfnamefont {E.}~\bibnamefont {Aschenauer}}, \bibinfo
  {author} {\bibfnamefont {G.}~\bibnamefont {Atoian}}, \bibinfo {author}
  {\bibfnamefont {K.~O.}\ \bibnamefont {Eyser}}, \bibinfo {author}
  {\bibfnamefont {H.}~\bibnamefont {Huang}}, \bibinfo {author} {\bibfnamefont
  {Y.}~\bibnamefont {Makdisi}}, \bibinfo {author} {\bibfnamefont {W.~B.}\
  \bibnamefont {Schmidke}}, \bibinfo {author} {\bibfnamefont {I.}~\bibnamefont
  {Alekseev}}, \bibinfo {author} {\bibfnamefont {D.}~\bibnamefont {Svirida}},\
  and\ \bibinfo {author} {\bibfnamefont {N.~H.}\ \bibnamefont {Buttimore}},\
  }\bibfield  {title} {\bibinfo {title} {{Precision Small Scattering Angle
  Measurements of Elastic Proton-Proton Single and Double Spin Analyzing Powers
  at the RHIC Hydrogen Jet Polarimeter}},\ }\href
  {https://doi.org/10.1103/PhysRevLett.123.162001} {\bibfield  {journal}
  {\bibinfo  {journal} {Phys. Rev. Lett.}\ }\textbf {\bibinfo {volume} {123}},\
  \bibinfo {pages} {162001} (\bibinfo {year} {2019})},\ \Eprint
  {https://arxiv.org/abs/1909.11135} {arXiv:1909.11135 [hep-ex]} \BibitemShut
  {NoStop}%
\bibitem [{\citenamefont {Poblaguev}(2022{\natexlab{a}})}]{Poblaguev:2022yzw}%
  \BibitemOpen
  \bibfield  {author} {\bibinfo {author} {\bibfnamefont {A.~A.}\ \bibnamefont
  {Poblaguev}},\ }\bibfield  {title} {\bibinfo {title} {{Feasibility study for
  precisely measuring the EIC ${}^3$He beam polarization with the Polarized
  Atomic Hydrogen Gas Jet Target polarimeter at RHIC}},\ }\href
  {https://doi.org/10.1103/PhysRevC.106.065202} {\bibfield  {journal} {\bibinfo
   {journal} {Phys. Rev. C}\ }\textbf {\bibinfo {volume} {106}},\ \bibinfo
  {pages} {065202} (\bibinfo {year} {2022}{\natexlab{a}})},\ \Eprint
  {https://arxiv.org/abs/2207.09420} {arXiv:2207.09420 [hep-ph]} \BibitemShut
  {NoStop}%
\bibitem [{\citenamefont {Kopeliovich}\ and\ \citenamefont
  {Lapidus}(1974)}]{Kopeliovich:1974ee}%
  \BibitemOpen
  \bibfield  {author} {\bibinfo {author} {\bibfnamefont {B.~Z.}\ \bibnamefont
  {Kopeliovich}}\ and\ \bibinfo {author} {\bibfnamefont {L.~I.}\ \bibnamefont
  {Lapidus}},\ }\bibfield  {title} {\bibinfo {title} {{On the necessity of
  polarization experiments in colliding $pp$ and $\overline{p}p$ beams}},\
  }\href@noop {} {\bibfield  {journal} {\bibinfo  {journal} {Yad. Fiz.}\
  }\textbf {\bibinfo {volume} {19}},\ \bibinfo {pages} {218} (\bibinfo {year}
  {1974})},\ \bibinfo {note} {{[Sov. J. Nucl. Phys. {\bf19}, 114
  (1974)]}}\BibitemShut {NoStop}%
\bibitem [{\citenamefont {Buttimore}(2009)}]{Buttimore:2009zz}%
  \BibitemOpen
  \bibfield  {author} {\bibinfo {author} {\bibfnamefont {N.~H.}\ \bibnamefont
  {Buttimore}},\ }\bibfield  {title} {\bibinfo {title} {{Forward helion
  scattering and neutron polarization}},\ }\href
  {https://doi.org/10.1063/1.3122170} {\bibfield  {journal} {\bibinfo
  {journal} {AIP Conf. Proc.}\ }\textbf {\bibinfo {volume} {1105}},\ \bibinfo
  {pages} {189} (\bibinfo {year} {2009})}\BibitemShut {NoStop}%
\bibitem [{\citenamefont {Kopeliovich}\ and\ \citenamefont
  {Trueman}(2001)}]{Kopeliovich:2000kz}%
  \BibitemOpen
  \bibfield  {author} {\bibinfo {author} {\bibfnamefont {B.~Z.}\ \bibnamefont
  {Kopeliovich}}\ and\ \bibinfo {author} {\bibfnamefont {T.~L.}\ \bibnamefont
  {Trueman}},\ }\bibfield  {title} {\bibinfo {title} {{Polarized proton nucleus
  scattering}},\ }\href {https://doi.org/10.1103/PhysRevD.64.034004} {\bibfield
   {journal} {\bibinfo  {journal} {Phys. Rev. D}\ }\textbf {\bibinfo {volume}
  {64}},\ \bibinfo {pages} {034004} (\bibinfo {year} {2001})},\ \Eprint
  {https://arxiv.org/abs/hep-ph/0012091} {arXiv:hep-ph/0012091} \BibitemShut
  {NoStop}%
\bibitem [{\citenamefont {Igo}\ and\ \citenamefont
  {Tanihata}(2003)}]{Igo:2003cs}%
  \BibitemOpen
  \bibfield  {author} {\bibinfo {author} {\bibfnamefont {G.}~\bibnamefont
  {Igo}}\ and\ \bibinfo {author} {\bibfnamefont {I.}~\bibnamefont {Tanihata}},\
  }\bibfield  {title} {\bibinfo {title} {{Absolute calibration of the RHIC CNI
  polarimeters using 125-GeV/A C ions}},\ }\href
  {https://doi.org/10.1063/1.1607251} {\bibfield  {journal} {\bibinfo
  {journal} {AIP Conf. Proc.}\ }\textbf {\bibinfo {volume} {675}},\ \bibinfo
  {pages} {836} (\bibinfo {year} {2003})}\BibitemShut {NoStop}%
\bibitem [{\citenamefont {Bellettini}\ \emph {et~al.}(1966)\citenamefont
  {Bellettini}, \citenamefont {Cocconi}, \citenamefont {Diddens}, \citenamefont
  {Lillethun}, \citenamefont {Matthiae}, \citenamefont {Scanlon},\ and\
  \citenamefont {Wetherell}}]{Bellettini:1966zz}%
  \BibitemOpen
  \bibfield  {author} {\bibinfo {author} {\bibfnamefont {G.}~\bibnamefont
  {Bellettini}}, \bibinfo {author} {\bibfnamefont {G.}~\bibnamefont {Cocconi}},
  \bibinfo {author} {\bibfnamefont {A.}~\bibnamefont {Diddens}}, \bibinfo
  {author} {\bibfnamefont {E.}~\bibnamefont {Lillethun}}, \bibinfo {author}
  {\bibfnamefont {G.}~\bibnamefont {Matthiae}}, \bibinfo {author}
  {\bibfnamefont {J.}~\bibnamefont {Scanlon}},\ and\ \bibinfo {author}
  {\bibfnamefont {A.}~\bibnamefont {Wetherell}},\ }\bibfield  {title} {\bibinfo
  {title} {{Proton-nuclei cross sections at 20 GeV}},\ }\href
  {https://doi.org/10.1016/0029-5582(66)90267-7} {\bibfield  {journal}
  {\bibinfo  {journal} {Nucl. Phys.}\ }\textbf {\bibinfo {volume} {79}},\
  \bibinfo {pages} {609} (\bibinfo {year} {1966})}\BibitemShut {NoStop}%
\bibitem [{\citenamefont {Glauber}\ and\ \citenamefont
  {Matthiae}(1970)}]{Glauber:1970jm}%
  \BibitemOpen
  \bibfield  {author} {\bibinfo {author} {\bibfnamefont {R.}~\bibnamefont
  {Glauber}}\ and\ \bibinfo {author} {\bibfnamefont {G.}~\bibnamefont
  {Matthiae}},\ }\bibfield  {title} {\bibinfo {title} {{High-energy scattering
  of protons by nuclei}},\ }\href
  {https://doi.org/10.1016/0550-3213(70)90511-0} {\bibfield  {journal}
  {\bibinfo  {journal} {Nucl. Phys. B}\ }\textbf {\bibinfo {volume} {21}},\
  \bibinfo {pages} {135} (\bibinfo {year} {1970})}\BibitemShut {NoStop}%
\bibitem [{\citenamefont {Liu}\ \emph {et~al.}(2017)\citenamefont {Liu} \emph
  {et~al.}}]{Liu:IPAC2017-TUPVA046}%
  \BibitemOpen
  \bibfield  {author} {\bibinfo {author} {\bibfnamefont {C.}~\bibnamefont
  {Liu}} \emph {et~al.},\ }\bibfield  {title} {\bibinfo {title} {{B}eam
  {E}nergy {S}can {W}ith {A}symmetric {C}ollision at {RHIC}},\ }in\ \href
  {https://doi.org/https://doi.org/10.18429/JACoW-IPAC2017-TUPVA046} {\emph
  {\bibinfo {booktitle} {Proc. of the 8th International Particle Accelerator
  Conference (IPAC'17), Copenhagen, Denmark, 14--19 May, 2017}}}\ (\bibinfo
  {publisher} {JACoW},\ \bibinfo {address} {Geneva, Switzerland},\ \bibinfo
  {year} {2017})\ pp.\ \bibinfo {pages} {2175--2177}\BibitemShut {NoStop}%
\bibitem [{\citenamefont {Purcell}\ and\ \citenamefont
  {Sheu}(2015)}]{Purcell:2015gtm}%
  \BibitemOpen
  \bibfield  {author} {\bibinfo {author} {\bibfnamefont {J.}~\bibnamefont
  {Purcell}}\ and\ \bibinfo {author} {\bibfnamefont {C.}~\bibnamefont {Sheu}},\
  }\bibfield  {title} {\bibinfo {title} {{Nuclear Data Sheets for A = 3}},\
  }\href {https://doi.org/10.1016/j.nds.2015.11.001} {\bibfield  {journal}
  {\bibinfo  {journal} {Nucl. Data Sheets}\ }\textbf {\bibinfo {volume}
  {130}},\ \bibinfo {pages} {1} (\bibinfo {year} {2015})}\BibitemShut {NoStop}%
\bibitem [{\citenamefont {Berestetskii}\ \emph {et~al.}(1982)\citenamefont
  {Berestetskii}, \citenamefont {Lifshitz},\ and\ \citenamefont
  {Pitaevskii}}]{Berestetsky:1982aq}%
  \BibitemOpen
  \bibfield  {author} {\bibinfo {author} {\bibfnamefont {V.}~\bibnamefont
  {Berestetskii}}, \bibinfo {author} {\bibfnamefont {E.}~\bibnamefont
  {Lifshitz}},\ and\ \bibinfo {author} {\bibfnamefont {L.}~\bibnamefont
  {Pitaevskii}},\ }\href@noop {} {\emph {\bibinfo {title} {{\em Quantum
  Electrodynamics}}}},\ Course of Theoretical Physics Vol. 4\ (\bibinfo
  {publisher} {Pergamon},\ \bibinfo {address} {Oxford},\ \bibinfo {year}
  {1982})\BibitemShut {NoStop}%
\bibitem [{\citenamefont {Huang}\ and\ \citenamefont
  {Zhou}(2005)}]{HUANG2005283}%
  \BibitemOpen
  \bibfield  {author} {\bibinfo {author} {\bibfnamefont {X.}~\bibnamefont
  {Huang}}\ and\ \bibinfo {author} {\bibfnamefont {C.}~\bibnamefont {Zhou}},\
  }\bibfield  {title} {\bibinfo {title} {Nuclear data sheets for a=197},\
  }\href {https://doi.org/https://doi.org/10.1016/j.nds.2005.01.001} {\bibfield
   {journal} {\bibinfo  {journal} {Nuclear Data Sheets}\ }\textbf {\bibinfo
  {volume} {104}},\ \bibinfo {pages} {283 } (\bibinfo {year}
  {2005})}\BibitemShut {NoStop}%
\bibitem [{\citenamefont {Atoian}\ \emph {et~al.}(2021)\citenamefont {Atoian},
  \citenamefont {Poblaguev},\ and\ \citenamefont {Zelenski}}]{Au_Breakup}%
  \BibitemOpen
  \bibfield  {author} {\bibinfo {author} {\bibfnamefont {G.}~\bibnamefont
  {Atoian}}, \bibinfo {author} {\bibfnamefont {A.~A.}\ \bibnamefont
  {Poblaguev}},\ and\ \bibinfo {author} {\bibfnamefont {A.}~\bibnamefont
  {Zelenski}},\ }\bibfield  {title} {\bibinfo {title} {{Experimental evaluation
  of the breakup rate in the $p\text{Au}$ scattering at HJET}}} (\bibinfo
  {year} {2021})\BibitemShut {NoStop}%
\bibitem [{\citenamefont {Shindin}\ \emph {et~al.}()\citenamefont {Shindin},
  \citenamefont {Guriev}, \citenamefont {Livanov},\ and\ \citenamefont
  {Yudin}}]{Shindin:2017mcs}%
  \BibitemOpen
  \bibfield  {author} {\bibinfo {author} {\bibfnamefont {R.~A.}\ \bibnamefont
  {Shindin}}, \bibinfo {author} {\bibfnamefont {D.~K.}\ \bibnamefont {Guriev}},
  \bibinfo {author} {\bibfnamefont {A.~N.}\ \bibnamefont {Livanov}},\ and\
  \bibinfo {author} {\bibfnamefont {I.~P.}\ \bibnamefont {Yudin}},\ }\bibfield
  {title} {\bibinfo {title} {{Interesting effect of the $nd\rightarrow p(nn)$
  reaction (in Russian)}},\ }\href@noop {} {\ }\Eprint
  {https://arxiv.org/abs/1703.08820} {arXiv:1703.08820 [nucl-th]} \BibitemShut
  {NoStop}%
\bibitem [{\citenamefont {Cahn}(1982)}]{Cahn:1982nr}%
  \BibitemOpen
  \bibfield  {author} {\bibinfo {author} {\bibfnamefont {R.}~\bibnamefont
  {Cahn}},\ }\bibfield  {title} {\bibinfo {title} {{Coulombic-Hadronic
  Interference in an Eikonal Model}},\ }\href
  {https://doi.org/10.1007/BF01475009} {\bibfield  {journal} {\bibinfo
  {journal} {Z. Phys. C}\ }\textbf {\bibinfo {volume} {15}},\ \bibinfo {pages}
  {253} (\bibinfo {year} {1982})}\BibitemShut {NoStop}%
\bibitem [{\citenamefont {Kopeliovich}\ and\ \citenamefont
  {Tarasov}(2001)}]{Kopeliovich:2000ez}%
  \BibitemOpen
  \bibfield  {author} {\bibinfo {author} {\bibfnamefont {B.~Z.}\ \bibnamefont
  {Kopeliovich}}\ and\ \bibinfo {author} {\bibfnamefont {A.~V.}\ \bibnamefont
  {Tarasov}},\ }\bibfield  {title} {\bibinfo {title} {{The Coulomb phase
  revisited}},\ }\href {https://doi.org/10.1016/S0370-2693(00)01316-2}
  {\bibfield  {journal} {\bibinfo  {journal} {Phys. Lett. B}\ }\textbf
  {\bibinfo {volume} {497}},\ \bibinfo {pages} {44} (\bibinfo {year} {2001})},\
  \Eprint {https://arxiv.org/abs/hep-ph/0010062} {arXiv:hep-ph/0010062}
  \BibitemShut {NoStop}%
\bibitem [{\citenamefont {Poblaguev}(2022{\natexlab{b}})}]{Poblaguev:2021xkd}%
  \BibitemOpen
  \bibfield  {author} {\bibinfo {author} {\bibfnamefont {A.~A.}\ \bibnamefont
  {Poblaguev}},\ }\bibfield  {title} {\bibinfo {title} {{Coulomb phase
  corrections to the transverse analyzing power $A_N(t)$ in high energy forward
  proton-proton scattering}},\ }\href
  {https://doi.org/10.1103/PhysRevD.105.096039} {\bibfield  {journal} {\bibinfo
   {journal} {Phys. Rev. D}\ }\textbf {\bibinfo {volume} {105}},\ \bibinfo
  {pages} {096039} (\bibinfo {year} {2022}{\natexlab{b}})},\ \Eprint
  {https://arxiv.org/abs/2111.01696} {arXiv:2111.01696 [hep-ph]} \BibitemShut
  {NoStop}%
\bibitem [{\citenamefont {Hulth\'{e}n}\ and\ \citenamefont
  {Sugawara}(1957)}]{Hulthen:1957}%
  \BibitemOpen
  \bibfield  {author} {\bibinfo {author} {\bibfnamefont {L.}~\bibnamefont
  {Hulth\'{e}n}}\ and\ \bibinfo {author} {\bibfnamefont {M.}~\bibnamefont
  {Sugawara}},\ }\bibfield  {title} {\bibinfo {title} {{The Two-Nucleon
  Problem}},\ }in\ \href
  {https://doi.org/https://doi.org/10.1007/978-3-642-45872-9_1} {\emph
  {\bibinfo {booktitle} {Structure of Atomic Nuclei}}},\ \bibinfo {series}
  {{Encyclopedia of Physics}}, Vol.~\bibinfo {volume} {39},\ \bibinfo {editor}
  {edited by\ \bibinfo {editor} {\bibfnamefont {S.}~\bibnamefont {Flugge}}}\
  (\bibinfo  {publisher} {Springer, Berlin, Heidelberg},\ \bibinfo {year}
  {1957})\ pp.\ \bibinfo {pages} {32--33, 76, 92}\BibitemShut {NoStop}%
\bibitem [{\citenamefont {Atanasov}\ \emph {et~al.}(1994)\citenamefont
  {Atanasov} \emph {et~al.}}]{Atanasov:1994}%
  \BibitemOpen
  \bibfield  {author} {\bibinfo {author} {\bibfnamefont {I.}~\bibnamefont
  {Atanasov}} \emph {et~al.},\ }\bibfield  {title} {\bibinfo {title} {{The
  measurements of the polarization transfer coefficient in $(d,p)$ reaction at
  fixed proton momentum of 4.5 GeV/c and the deuteron momentum in range
  6.0--9.0 GeV/c}},\ }in\ \href@noop {} {\emph {\bibinfo {booktitle} {Proc. of
  the 11th International Baldin Seminar on High Energy Physics Problems :
  Relativistic Nuclear Physics and Quantum Chromodynamics (ISHEPP 1992), Dubna,
  7--12 September, 1992}}},\ \bibinfo {editor} {edited by\ \bibinfo {editor}
  {\bibfnamefont {A.~M.}\ \bibnamefont {Baldin}}\ and\ \bibinfo {editor}
  {\bibfnamefont {V.~V.}\ \bibnamefont {Burov}}}\ (\bibinfo  {publisher} {JINR
  Dubna},\ \bibinfo {year} {1994})\ p.\ \bibinfo {pages} {443}\BibitemShut
  {NoStop}%
\end{thebibliography}
\end{document}